\documentclass[MSNbibl,nameyear,dvips]{arxstspdf}
\usepackage{dcolumn}
\usepackage{graphicx}
\usepackage{flushend}
\usepackage{stfloats}

\volume{28}
\issue{3}
\pubyear{2013}
\firstpage{398}
\lastpage{423}
\doi{10.1214/13-STS433} 

\makeatletter

\newcolumntype{d}[1]{D{.}{.}{#1}}

\newcommand{\rrVert}{\Vert}
\newcommand{\rrvert}{\vert}
\newcommand{\llVert}{\Vert}
\newcommand{\llvert}{\vert}

\newcommand{\cal}{\mathcal}

\newproclaim{defin}{Definition}[section]
\newproclaim{example}{Example}[section]
\newtheorem{lemma}{Lemma}[section]
\newtheorem{theorem}{Theorem}[section]
\newtheorem{cor}{Corollary}[section]
\newproclaim{result}{Result}[section]

\newcommand{\prob}{\mathbb{P}}
\newcommand{\vat}{\mathbb{E}}
\newcommand{\var}{\mathbb{V}}
\newcommand{\g}{|}

\makeatother

\begin{document}
\begin{frontmatter}

\title{The Whetstone and the Alum Block: Balanced Objective Bayesian
Comparison of
Nested Models for Discrete Data}
\runtitle{Objective Bayesian Model Comparison}

\begin{aug}
\author[a]{\fnms{Guido} \snm{Consonni}\corref{}\ead[label=e1]{guido.consonni@unicatt.it}},
\author[b]{\fnms{Jonathan J.} \snm{Forster}\ead[label=e2]{j.j.forster@soton.ac.uk}}
\and
\author[c]{\fnms{Luca} \snm{La Rocca}\ead[label=e3]{luca.larocca@unimore.it}}
\runauthor{G. Consonni, J. J. Forster and L. La Rocca}

\affiliation{Universit\`a Cattolica del Sacro Cuore, University of
Southampton and
Universit\`a di Modena e Reggio Emilia}

\address[a]{Guido Consonni is Professor of Statistics,
Dipartimento di Scienze Statistiche, Universit\`a Cattolica del Sacro Cuore,
Milano, Italy \printead{e1}.}
\address[b]{Jonathan J. Forster is Professor of Statistics, School of
Mathematics and
Southampton Statistical Sciences Research Institute, University of Southampton,
Southampton SO17 1BJ,
United Kingdom \printead{e2}.}
\address[c]{Luca La Rocca is Assistant Professor of Statistics,
Dipartimento di Scienze Fisiche, Informatiche e Matematiche,
Universit\`a di Modena e Reggio Emilia,
Modena, Italy \printead{e3}.}

\end{aug}

%
\begin{abstract}
When two nested models are compared, using a Bayes factor,
from an objective standpoint, two seemingly conflicting issues emerge
at the time of choosing parameter priors under the two models.
On the one hand, for moderate sample sizes, the evidence in favor of
the smaller model
can be inflated by diffuseness of the prior under the larger model.
On the other hand, asymptotically, the evidence
in favor of the smaller model typically accumulates at a slower rate.
With reference to finitely discrete data models, we show that these two issues
can be dealt with jointly,
by combining intrinsic priors and nonlocal priors in
a new unified class of priors.
We illustrate our ideas in a running Bernoulli example,
then we apply them to test the equality of two proportions,
and finally we deal with the more general case of logistic regression models.
\end{abstract}

%
\begin{keyword}
\kwd{Bayes factor}
\kwd{intrinsic prior}
\kwd{model choice}
\kwd{moment prior}
\kwd{nonlocal prior}
\kwd{Ockham's razor}
\kwd{training sample size}
\end{keyword}

\end{frontmatter}

\section{Introduction}\label{sec1}

Consider two parametric models,
${\cal M}_0$ (the \emph{null} model)
nested in ${\cal M}_1$ (the \emph{alternative} model),
each equipped with its own prior distribution,
$p_0(\cdot)$ and $p_1(\cdot)$.
We plan to compare models using the Bayes Factor (BF);
see \citet{KassRaft1995} for a classic review.
We denote by $f_i(\cdot|\theta_i)$ the sampling density of data $y$
under ${\cal M}_i$,
$i=0,1$. Then, the BF in favor of ${\cal M}_1$, or against ${\cal M}_0$,
is defined as $\mathrm{BF}_{10}(y)=m_1(y)/m_0(y)$,
where $m_i(y)= \int f_i(y|\theta_i)p_i(\theta_i)\,d\theta_i$ is the
marginal density of $y$
under ${\cal M}_i$, also called the marginal likelihood of ${\cal M}_i$.

It is well known that special care must be exercised in the
specification of $p_0(\cdot)$
and $p_1(\cdot)$ when computing the BF. One obvious condition is that
neither prior
be improper, because the resulting BF would depend on arbitrary
constants.
Even when proper priors are used, however, difficulties may arise;
in particular, this happens when $p_1(\cdot)$ is not chosen in view of
the comparison
with model ${\cal M}_0$, which is of course the rule with conventional priors.
In general, a conventional $p_1(\cdot)$ will be rather diffuse,
and will thus give little weight to sampling densities close to
the subspace characterizing~${\cal M}_0$.
Therefore, unless the data are vastly against~${\cal M}_0$,
which rarely happens for moderate sample sizes,
there will be an evidence bias in favor of ${\cal M}_0$.
Informally, this happens because $p_1(\cdot)$
``wastes'' probability mass in parameter areas too remote from the null.
This fact had essentially been realized as early as in
\citeauthor{Jeff1961} (\citeyear{Jeff1961}, Chapter 3)
and was already clear in \citet{Morr1987}, whose suggestion was
to ``center''
$p_1(\cdot)$ around the null-subspace.
In this spirit, we will argue in favor of ``transferring probability mass''
toward the null subspace within a given diffuse prior under ${\cal M}_1$.

Although we used no limiting argument above,
there is a connection with the Jeffreys--Lindley--Bart\-lett paradox;
see
\citeauthor{OhagFors2004} (\citeyear{OhagFors2004},
Section~3.33), \citet{RobeChopRous2009} and \citet{Senn09}.
According to one version of the paradox, if the sample size is fixed,
but the variance of $p_1(\cdot)$ is free to increase without bound,
the posterior probability of the null model will go to one,
irrespective of the data. This is just an exacerbation of the phenomenon
described above, with $p_1(\cdot)$ allocating probability mass in unreasonable
regions of the parameter space.

A word of caution is useful at this stage. From a Bayesian perspective,
a model is a \emph{pair}, whose elements are the family of sampling
distributions
(sampling model) and the prior.
Nevertheless, we will follow the prevailing practice of using the word ``model''
to identify the sampling model, leaving to the prior the role of specifying
which Bayesian model is actually entertained.

Adhering to an objective viewpoint, we assume that default parameter priors
$p_i(\cdot)$, $i=1,2$, are given, each of them depending only on the
corresponding model.
We also assume, for simplicity, that both priors are proper.
The action of reallocating mass within $p_1(\cdot)$ toward the null subspace
has a negative side effect,
at least for moderate sample sizes: it will diminish evidence in favor
of ${\cal M}_1$
when the parameter values generating the data are truly away from the null.
However, this price is worth paying, to some extent, because of two reasons:
(i) the very fact that we are considering ${\cal M}_0$ testifies that
it has some
a priori plausibility and, thus, parameter values close to the
null are
more interesting to monitor than those remote from it; (ii) if the data
manifestly support
${\cal M}_1$, we can surely afford the luxury to somewhat diminish the strength
of evidence in its favor, because it will be already high enough for
most practical purposes.
However, it is not at all obvious how far this strategy should be pushed,
and we dedicate part of this paper to try and answer this question.

In light of the above discussion, two general issues are to be addressed
in the setting under consideration:
\begin{longlist}[(2)]
\item[(1)]
given model ${\cal M}_0$ nested in ${\cal M}_1$,
and the corresponding default priors $p_0(\cdot)$ and $p_1(\cdot)$,
how can we build an ${\cal M}_0$-focused prior under ${\cal M}_1$,
transferring probability mass within $p_1(\cdot)$ toward the null subspace
characterizing ${\cal M}_0$?
\item[(2)]
how do we settle the \emph{evidence trade-off}:
reinforcing the evidence in favor of ${\cal M}_1$ for parameter values
around the null subspace, while weakening it when the parameter lies
in regions away from the null?
\end{longlist}

A possible objection is that point (1) could be bypassed:
once we have understood the features of a ``good'' parameter prior
under ${\cal M}_1$,
why should we bother with the default prior anyway? We accept this criticism
and do not object to a subjective specification carefully taking into account
the \emph{desiderata} we set out.
We remark, however, that this task may be far from simple,
requires substantive knowledge not always available,
and could become daunting when many pairwise comparisons
are entertained (like in variable selection).
This is the reason why we privilege an objective approach,
which takes as input only the default priors.


A natural answer to point (1) is provided by the \emph{intrinsic priors},
whose scope is indeed not restricted to nested models.
Intrinsic priors are now recognized as an important tool
for objective Bayesian hypothesis testing and model comparison.
Numerous applications witness their usefulness,
ranging from variable selection
(\cite{CaseMore2006}; \cite{CaseEtAl2009};
\cite{MoreEtal2010}; \cite{LeonEtal2012})
to contingency tables
(Casella and Moreno \citeyear{CaseMore2005,CaseMore2009}; \cite{ConsLaRo2008};
\cite{ConsMoreVent2011})
to change point problems
(\cite{MoreEtAl2005}; \cite{GiroEtAl2007}).
When the two models are nested, the end result of the intrinsic prior procedure
is to modify $p_1(\cdot)$ so that the resulting intrinsic prior
accumulates more mass
around the null subspace. This is achieved by mixing over a \emph
{training sample},
whose size $t$ regulates the amount of concentration of the intrinsic prior,
which we denote by $p^I_1(\cdot|t)$, around the null subspace.

If $p_1(\cdot)$ is improper,
it is tempting to set $t$ in the intrinsic prior $p^I_1(\cdot|t)$
equal to the \emph{minimal} training sample size,
which is the smallest sample size for which the default posterior is
proper on all data.
However, no formal justification is available for this choice,
which clearly bypasses point (2) on grounds of simplicity.
On the other hand,\vadjust{\goodbreak} when the default prior is proper,
as happens in some discrete data problems, there is no general
guideline for fixing $t$,
and usually a robustness analysis is performed by letting $t$ vary
between $1$ and $n$,
where $n$ is the actual sample size; see, for instance, \citet
{CaseMore2009}.
Point (2) is here bypassed in favor of a sensitivity analysis.

Intrinsic priors for the comparison of nested models can be viewed as
\emph{expected posterior priors} (\cite{PereBerg2002}) with baseline
mixing distribution
equal to the marginal data distribution under ${\cal M}_0$.
Another related approach is due to \citet{Neal2001}:
since subjective prior elicitation of the parameter prior should be
more precise,
and possibly easier, under the smaller model than under the larger model,
information should be transferred from the former to the latter
by means of a training sample,
whose sample size $t$ will determine how similar or compatible the two models
turn out to be. Neal offers no guidance on fixing $t$;
interestingly, however, in his approach $t$ can grow to infinity.

An alternative to the intrinsic (or expected posterior) prior approach
to derive the BF in the presence of improper priors is the
Fractional Bayes Factor (FBF); see \citet{Ohag1995}.
Here a fraction $b$ of the likelihood is used to obtain a fractional
posterior distribution,
which in turn is used as a (data-dependent) prior to construct a
fractional marginal likelihood based on the likelihood raised to the
complementary fraction $(1-b)$. This calculation is repeated under both models.
The end result is to shift the prior under each of the two models
toward a region supported by the likelihood.
Clearly, if the data are in reasonable accord with ${\cal M}_0$, which
we have identified
as the most critical situation for nested model comparison,
the prior under ${\cal M}_1$ will tend to concentrate around the null subspace,
like in the intrinsic prior approach. In this sense, the fraction $b$
plays a role
akin to that of the training sample size $t$, although it should be
stressed that
the implied prior in the FBF is data dependent, while this is not the
case for
the intrinsic prior. A conventional choice is $b=n_0/n$, where $n_0$ is
the smallest
integer that makes the fractional posteriors proper.
\citeauthor{Ohag1995} (\citeyear{Ohag1995}, Section 6)
also suggests two alternative choices for cases when robustness is a major
concern, but \citet{More1997} has an argument against these choices.
Data-centered priors for each of the two models can also be constructed using
the expected posterior priors of \citet{PereBerg2002}, setting the
baseline mixing distribution equal to the empirical distribution.\vadjust{\goodbreak}

The FBF is typically easier to implement than the intrinsic approach.
However, the fact that it uses a data dependent prior is clearly a drawback.
Accordingly, since implementation issues turn out to be less compelling
for discrete data problems, in this paper we address point (1) through
intrinsic priors.
As for the issue raised in point (2),
which to the best of our knowledge has never been tackled so far,
we propose a solution in Section~\ref{subsecChoosingTrainingSampleSize}
based on the notion of \emph{total weight of evidence}.
We do not claim that our solution is universal,
but we found it useful in some examples,
and we believe that it sheds light on the evidence trade-off.


We now turn to a related aspect, which has been relatively neglected
in the literature on priors for model comparison: the asymmetry in the
learning rate
of the BF between two nested models $\mathcal{M}_0$ and $\mathcal{M}_1$.
A typical prior $p_1(\cdot)$, whether subjective or objective,
is continuous and strictly positive on the null subspace.
The second condition (given the first one) makes it a \emph{local} prior.
A serious deficiency of local priors relates to their asymptotic
learning rate.
Specifically, the BF in favor of $\mathcal{M}_1$,
when $\mathcal{M}_1$ holds, diverges in probability exponentially fast,
as the sample size grows; on the other hand, when $\mathcal{M}_0$ holds,
the same BF converges to zero in probability at a polynomial rate only.
Although this fact is well established,
it did not receive very much attention until \citet{JohnRoss2010}
brought it to the fore.
\citeauthor{RobeChopRous2009} (\citeyear{RobeChopRous2009}, Section 7)
report evidence that Jeffreys was aware of the
asymmetry, but that later studies neglected it. In practice,
the problem is that the imbalance is already quite dramatic
for moderate sample sizes. However, as suggested by \citet{JohnRoss2010},
this unsatisfactory feature can be corrected by using \emph{nonlocal} priors.
As the name suggests, these priors are built in opposition to local priors,
and their distinguishing feature, assuming continuity, is to be
identically zero
on the null subspace.

We find the idea of nonlocal priors appealing,
not only because they ameliorate the learning rate of the BF,
but also because they force the user to think more carefully
about the notion of \emph{model separation}.
This is a difficult issue, of course, which only occasionally can be answered
employing subject-matter knowledge; a notable example, reported in
\citet{Cohe1992},
is that standardized effect sizes of less than 0.2, in absolute value,
are often not considered substantively important in the social sciences.
However, we hasten to say that nonlocal priors have been disapproved by\vadjust{\goodbreak}
some authors;
see, for instance, the discussions of \citet{ConsLaro2011} by J.~Q.~Smith
and J. Rousseau with C. P. Robert.

Intrinsic priors and nonlocal priors play complementary roles in the
comparison of nested models. If the BF is seen as an implementation of
Ockham's razor,
the principle that an explanation (model)\break should not be more complicated
than necessary (\emph{pluralitas non est ponenda sine necessitate}),
as suggested by \citet{Jaynes1979}, \citet{SmitSpie1980},
and \citet{JeffBerg1992},
then Bayesian barbers should worry both about the sharpness of their tool,
on the one hand, and the risk of cutting the throat of the larger model,
on the other hand. Intrinsic priors protect the larger model from being
treated unfairly,
and thus play the role of an alum block, whereas nonlocal priors can greatly
sharpen the blade of the razor, and thus play the role of a whetstone.
A skilled combination of the two tools helps the Bayesian barber to
achieve a balanced comparison of the two models. In fact, we show in
this paper
that a suitably defined family of nonlocal intrinsic priors produces a
BF with
finite sample properties
comparable to those of ordinary (local) intrinsic priors,
and with the improved learning rate (when the null model holds)
characterizing nonlocal priors.

The rest of the paper is organized as follows.
Section~\ref{secpriorsTesting} provides background material on
intrinsic priors
and on a particular class of nonlocal priors, \emph{moment priors},
using as illustration
the problem of testing a sharp null hypothesis on a Bernoulli proportion.
Section~\ref{secintrimom} presents the class of \emph{intrinsic
moment priors},
for the comparison of two nested models, which is implemented in
Section~\ref{sectwoBinomial} for testing the equality of two proportions
and in Section~\ref{seclogreg} for variable selection in logistic
regression models.
Section~\ref{secappl} applies the suggested testing procedures
to a collection of randomized binary trials of a new surgical treatment
for stomach ulcers,
also discussed from a meta-analysis perspective by \citet{Efro1996},
and to a medical data set already analyzed by \citet{DellaEtal2002}
using logistic regression models.
Finally, Section~\ref{secdisc} offers some concluding remarks and investigates
a few issues worth further consideration.
A technical \hyperref[app]{Appendix} on the asymptotic learning rate of BFs completes
the paper.

\section{Priors for the Comparison of Nested Models}
\label{secpriorsTesting}

We review in this section two methodologies for constructing priors
when two nested models are compared: intrinsic priors and a specific class
of nonlocal priors, called moment priors.


Consider two sampling models for the same \emph{discrete} vector of
observables $y$:
%
\begin{eqnarray}
\label{samplingModelGeneral}
{\cal M}_0 &=& \bigl\{f_0(
\cdot|\xi_0), \xi_0 \in\Xi_0\bigr\}
\quad\mbox{vs.}\nonumber\\[-8pt]\\[-8pt]
{\cal M}_1&=&\bigl\{f_1(\cdot|
\xi_1), \xi_1 \in\Xi_1\bigr\},\nonumber
\end{eqnarray}
where ${\cal M}_0 $ is nested in ${\cal M}_1$, that is,
for all $\xi_0 \in\Xi_0$,
$f_0(\cdot|\xi_0)=f_1(\cdot|\xi_1)$, for\vspace*{1pt} some $\xi_1\in\tilde
{\Xi}_0\subset\Xi_1$,
where $\tilde{\Xi}_0$ is isomorphic to $\Xi_0$ and of lower
dimensionality than~$\Xi_1$.
Let $p_0(\cdot)$ be a given prior under ${\cal M}_0$
and similarly for $p_1(\cdot)$ under ${\cal M}_1$,
both of them being \emph{proper};
this assumption simplifies the exposition and is not particularly restrictive
because we deal with discrete data models.
Typically both $p_0(\cdot)$ and $p_1(\cdot)$ will be default,
inference-based, priors.
We also assume, again for simplicity,
equal prior probabilities for $\mathcal{M}_0$ and~$\mathcal{M}_1$,
so that the posterior probability of ${\cal M}_1 $ is a function of the
BF only:
$\prob(\mathcal{M}_1|y)=(1+\mathrm{BF}_{01}(y))^{-1}$,
where $\mathrm{BF}_{01}(y)=1/\mathrm{BF}_{10}(y)$ is the BF in favor of $\mathcal{M}_0$.

\subsection{Intrinsic Priors}

Intrinsic priors were introduced in objective hypothesis testing
to deal meaningfully with improper default priors when constructing BFs;
see \citet{BergPeri1996}, \citet{More1997}, Moreno, Berto\-lino and
Racugno (\citet{MoreEtAl1998}).
However, this view of the intrinsic prior approach is unduly restrictive
and actually hinders its inherent nature, as it is apparent
for discrete data models: in this case default priors are usually proper,
but the intrinsic approach can still be very useful.

As recalled in the \hyperref[sec1]{Introduction},
a default prior $p_1(\cdot)$ is typically inappropriate for testing purposes,
because it assigns little mass around the null subspace $\tilde{\Xi}_0$.
Mixing over the training sample $x=(x_1,\break\ldots, x_t)$,
the intrinsic prior on $\xi_1$ can be written as
%
\begin{equation}
\label{intriGeneral}\quad p^I_1(\xi_1|t)=\sum
_x p_1(\xi_1|x)m_0(x),\quad
\xi_1\in\Xi_1,
\end{equation}
where $p_1(\xi_1|x)$ is the posterior density of $\xi_1$ under~${\cal M}_1$,
given $x$, and $m_0(x)=\int f_0(x|\xi_0)p_0(\xi_0)\,{d}\xi_0$
is the marginal density of $x$ under $\mathcal{M}_0$;
it is natural to let $t=0$ in $p^I_1(\cdot|t)$ return the default
prior $p_1(\cdot)$.


We remark that (\ref{intriGeneral}) is not the original definition
of intrinsic prior, but rather its formulation as an expected posterior prior
(\cite{PereBerg2002}).
We find formula (\ref{intriGeneral}) especially appealing,
because it makes clear that an intrinsic prior is a mixture of
``posterior'' distributions. As we will illustrate shortly,
if the training sample size $t$ grows,
the intrinsic prior increases its concentration on the subspace $\tilde
{\Xi}_0$.
This is apparent from (\ref{intriGeneral}), because the weights $m_0(x)$
in the mixture will be higher for realizations $x$ more likely
under ${\cal M}_0$, and these realizations $x$ will drive the posterior
$p_1(\cdot|x)$ toward parameter values more supported under ${\cal M}_0$.
Notice that, if $t$ grows to infinity, the two Bayesian models
$(\{f_0(\cdot|\xi_0),\xi_0\in\Xi_0\},p_0(\cdot))$
and $(\{f_1(\cdot|\xi_1),\xi_1\in\Xi_1\},p_1(\cdot))$
will coincide, making the comparison problem trivial.
Section~\ref{subsecChoosingTrainingSampleSize} will discuss in greater detail
the nature of $t$ and will present a method to choose its value.

The BF based on the intrinsic prior is a weighted average of
conditional BFs
based on the default prior:
%
\begin{equation}
\label{eqintriGeneralBF} \mathrm{BF}_{10}^{I}(y|t)=\sum
_{x} 
\mathrm{BF}_{10}(y|x)m_0(x),
\end{equation}
where $\mathrm{BF}_{10}(y|x)$ is the BF obtained using $p_1(\cdot|x)$ as prior
under model ${\cal M}_1$; see, for example,
Consonni and La~Rocca (\citeyear{ConsLaRo2008}), Proposition 3.4.
Hence, at least for small $t$ and conjugate $p_1(\cdot)$,
computing $\mathrm{BF}_{10}^{I}(y|t)$ is not much more demanding than computing
$\mathrm{BF}_{10}(y)$.


\begin{example}[(Bernoulli)]
\label{exbernoulli}
Consider the testing problem
$\mathcal{M}_0\dvtx f_0(y|\theta_0)=\operatorname{Bin}(y|n,\theta_0)$
versus
$\mathcal{M}_1\dvtx\break f_1(y|\theta)=\operatorname{Bin}(y|n,\theta)$,
where $\theta_0$ is a \emph{fixed} value, while $\theta$ varies in $(0,1)$.
Let the default prior be $p_1(\theta|b)=\operatorname{Beta}(\theta|b,b)$
for some $b>0$.
We take a \emph{symmetric} prior because default objective priors
typically satisfy this property. In particular, letting $b=1/2$, we obtain
Jeffreys's prior, whereas $b=1$ gives us the uniform prior.
The intrinsic prior in this example is given by
%
\begin{eqnarray}
\label{eqintriPriorBern}
&&
p^{I}_1(\theta|b,t)\nonumber\\[-8pt]\\[-8pt]
&&\quad= \sum
_{x=0}^t \operatorname{Beta}(\theta|b+x,b+t-x)
\operatorname{Bin}(x|t,\theta_0).\nonumber
\end{eqnarray}
The solid curves in Figure~\ref{figbernoullifirst}(a), that is,
those specified by $h=0$, illustrate the shape of the intrinsic priors
with training sample size $t=0$ (default prior), $t=1$ and \mbox{$t=8$},
when $\theta_0=0.25$ and $b=1$.
The dashed curves ($h=1$) should be disregarded for the time being.
The effect of the intrinsic procedure is very clear:
already with $t=1$ the density has become a straight line with negative slope,
so as to start privileging low values of~$\theta$, such as $\theta_0=0.25$,
and with a training sample size \mbox{$t=8$} the effect is much more dramatic,
with the density now having a mode somewhere around $0.25$ and then
declining quickly.

\begin{figure*}

\includegraphics{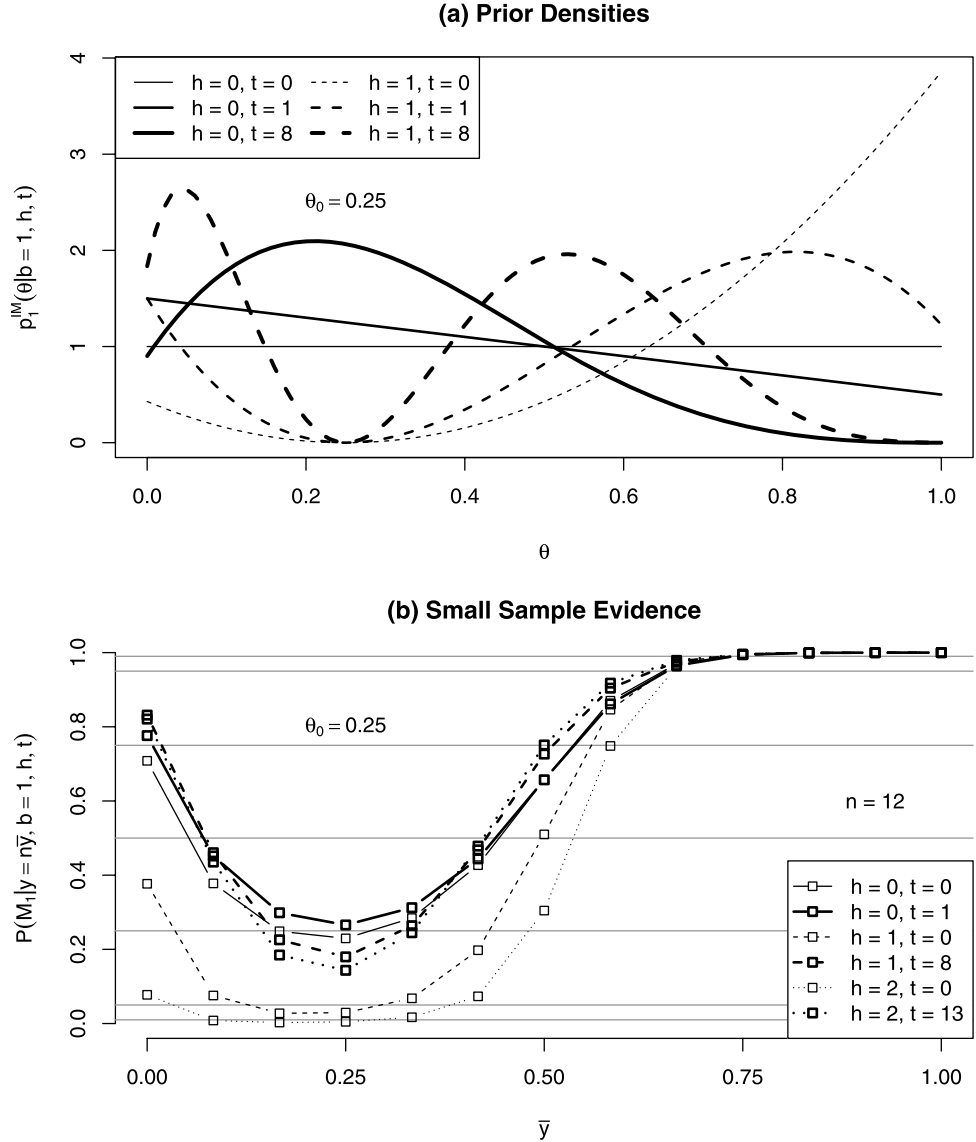}

\caption{Prior densities \textup{(a)} and small sample evidence \textup{(b)} for the
Bernoulli example. Horizontal gray lines in \textup{(b)} denote possible
decision thresholds at 1\%, 5\%, 25\%, 50\%, 75\%, 95\% and 99\% on the
posterior probability scale.} \label{figbernoullifirst}
\end{figure*}

Figure~\ref{figbernoullifirst}(b) shows (again focus on solid lines only)
the effect of the above-described probability mass transfer on the
comparison between $\mathcal{M}_0$ and $\mathcal{M}_1$:
for a small sample situation ($n=12$) the posterior probability of
$\mathcal{M}_1$,
computed from
%
\begin{eqnarray}
\label{eqintriPriorBernBF}
&&
\mathrm{BF}_{10}^{I}(y|b,t)\nonumber\\
&&\quad=\sum
_{x=0}^t 
\frac{B(b+x+y,b+t-x+n-y)}{B(b+x,b+t-x)\theta_0^y(1-\theta_0)^{n-y}}\\
&&\hspace*{14.7pt}\qquad{}\cdot
\operatorname{Bin}(x|t,\theta_0),\nonumber
\end{eqnarray}
where $B(\cdot,\cdot)$ denotes the Beta function,
is represented as a function of the observed frequency $\bar{y}$
(evidence curve)
both for the default prior and for the intrinsic prior with $t=1$.
The evidence curve reaches a minimum at $\bar{y}=0.25$
(data perfectly supporting the null)
and is somewhat higher for the intrinsic prior than for the
default prior when $0<\bar{y}<0.5$, because with the intrinsic prior
$\mathcal{M}_1$
becomes a stronger competitor when the data moderately support
$\mathcal{M}_0$.
\end{example}

Results similar to those given by the intrinsic prior can be obtained,
in the above example, by using a suitable Beta prior centered at
$\theta_0$.
However, the probability mass transfer toward $\theta_0$ takes place
more smoothly
under the intrinsic prior than under a Beta prior with increasing precision,
because the intrinsic prior is a mixture of Beta distributions.
Moreover, this kind of alternative approach is only available because
we are testing
a sharp hypothesis. When testing a composite hypothesis,
it is not at all obvious that a suitable conjugate prior can be found
(after a reparametrization of the model to identify a parameter
of interest and a nuisance parameter).
On the other hand, as we will see in Section~\ref{sectwoBinomial}
for the comparison of two proportions, the intrinsic prior produces
the desired outcome in a natural and automatic way.

\subsection{Moment Priors}
\label{subsecmoment}

Consider the testing problem (\ref{samplingModelGeneral}).
We say that \emph{the smaller model holds}
if the sampling distribution of the data belongs to $\mathcal{M}_0$;
we say that \emph{the larger model holds} if it belongs to $\mathcal{M}_1$
but not to $\mathcal{M}_0$. The following result shows an imbalance
in the learning rate of the BF for commonly used priors.

\begin{result}
In the testing problem (\ref{samplingModelGeneral})
assume that $p_0(\cdot)$ and $p_1(\cdot)$ are continuous
and strictly positive on $\Xi_0$ and $\Xi_1$, respectively, such that
some regularity conditions are satisfied by the two models
and that the data $y^{(n)}=(y_1,\ldots,y_n)$ arise\vspace*{1pt} under i.i.d. sampling.
If ${\cal M}_0$ holds,
then $\mathrm{BF}_{10}(y^{(n)})=n^{-(d_1-d_0)/2}e^{O_p(1)}$, as $n\to\infty$,
where $d_j$ is the dimension of $\Xi_j$, $j=1,2$, with $d_1>d_0$;
if ${\cal M}_1$ holds, then\vspace*{1pt} $\mathrm{BF}_{01}(y^{(n)})=e^{-Kn+O_p(n^{1/2})}$,
as $n\to\infty$, for some\break \mbox{$K>0$}.
\end{result}

We refer to \citet{Dawi2011} for a proof of this result.
It should be noted that a crucial role is played by the fact that
$p_1(\xi_1)>0$ for all $ \xi_1 \in\tilde{\Xi}_0$;
also recall that $p_1(\cdot)$ is continuous.
Thus, the only way to speed up the decrease of $\mathrm{BF}_{10}(y^{(n)})$,
when ${\cal M }_0$ holds, is to force the prior density under ${\cal M }_1$
to vanish on $\tilde{\Xi}_0$.


Let\vspace*{1pt} $g_h(\cdot)$ be a smooth function from $\Xi_1$ to $\Re_+$
vanishing on $\tilde{\Xi}_0$, together with its first $2h-1$ derivatives,
while $g_h^{(2h)}(\xi)$ is different from zero for all $\xi\in\tilde
{\Xi}_0$;
assume that $\int_{\Xi_1}g_h(\xi_1)p_1(\xi_1)\,d\xi_1$ is finite and nonzero.
Starting from a given local prior $p_1(\cdot)$,
we define the \emph{generalized} moment prior with moment function
$g_h(\cdot)$ as
%
\begin{equation}
\label{eqmomentPrior} p^{M}_1(\xi_1|h)
\propto g_h(\xi_1) p_1(\xi_1),\quad
\xi_1 \in \Xi_1.
\end{equation}
We impose that $g_0(\xi_1)\equiv1$,
so that setting $h=0$ in $p^M_1(\cdot|h)$
returns the local prior $p_1(\cdot)$.
For instance, if $\Xi_1 \subseteq\Re$
and $\tilde{\Xi}_0=\Xi_0=\{\xi_0\}$,
with $\xi_0$ a fixed value,
we may take $g_h(\xi_1)=(\xi_1-\xi_0)^{2h}$;
this defines the moment prior introduced by \citet{JohnRoss2010}
for testing a sharp hypothesis on a scalar parameter.
We refer to $h$ as the \emph{order} of the (generalized) moment prior.

The BF against ${\cal M}_0$ based on prior (\ref{eqmomentPrior})
can be computed as
%
\begin{eqnarray}
\label{eqmomentBF}
&&\mathrm{BF}^M_{10}\bigl(y^{(n)}|h
\bigr)\nonumber\\[-8pt]\\[-8pt]
&&\quad=\frac{\int_{\Xi_1}g_h(\xi_1) p_1(\xi
_1|y^{(n)})\,d\xi_1} {
\int_{\Xi_1}g_h(\xi_1)
p_1(\xi_1)\,d\xi_1}\mathrm{BF}_{10}\bigl(y^{(n)}\bigr),\nonumber
\end{eqnarray}
so that the extra effort required by using this prior amounts to
computing some (generalized) moments of the local prior and posterior.
This effort is rewarded by a reduction in the learning rate imbalance:
$\mathrm{BF}^M_{10}(y^{(n)}|h)=n^{-h-(d_1-d_0)/2}e^{O_p(1)}$,
when\break ${\cal M }_0$ holds, while we still have
$\mathrm{BF}^M_{01}(y^{(n)}|h)=\break e^{-Kn+O_p(n^{1/2})}$, when ${\cal M }_1$ holds;
see the \hyperref[app]{Appendix} for a justification of this result,
which generalizes the rates found by \citet{JohnRoss2010}
for their specific moment priors.

\begin{example}[(Bernoulli c.t.d.)]
\label{exbernoullictd}
Starting from the local prior $\operatorname{Beta}(\theta|a_1,a_2)$,
we define the moment prior of order $h$ as
%
\begin{eqnarray}
\label{MC}
&&
p^M_1(\theta|a_1,a_2,h)\nonumber\\[-8pt]\\[-8pt]
&&\quad=
\frac{(\theta-\theta_0)^{2h}} {
K(a_1,a_2,h,\theta_0)} \operatorname{Beta}(\theta|a_1,a_2),\nonumber
\end{eqnarray}
where
%
\begin{eqnarray}
\label{Kgeneral}
&&
K(a_1,a_2,h,\theta_0)\nonumber\\[-8pt]\\[-8pt]
&&\quad=
\frac{\theta_0^{2h}}{B(a_1,a_2)}\sum_{j=0}^{2h} \pmatrix{2h
\cr
j}(-1)^j\theta_0^{-j} B(a_1+j,a_2),\hspace*{-15pt}\nonumber
\end{eqnarray}
%
and obtain
%
\begin{eqnarray}
\label{eqmomentBFbern}
&&
\mathrm{BF}^M_{10}(y|a_1,a_2,h)\nonumber\\
&&\quad=
\frac{K(a_1+y,a_2+n-y,h,\theta _0)}{K(a_1,a_2,h,\theta_0)}\\
&&\qquad{}\cdot
\frac{B(a_1+y,a_2+n-y)}{B(a_1,a_2)\theta_0^y(1-\theta_0)^{n-y}}.\nonumber
\end{eqnarray}
In particular, we are interested in the default choice $a_1=a_2=b$
(with $b=1/2$ or $b=1$). The moment prior $p^M_1(\theta|b,h)$
is represented in Figure~\ref{figbernoullifirst}(a),
for \mbox{$b=1$} and $\theta_0=0.25$, by the thin dashed curve specified by
$h=1$ and $t=0$.\vadjust{\goodbreak}
The other two dashed curves, specified by $h=1$ and $t=1$ (intermediate curve)
or $t=8$ (thick curve), should be ignored for the time being.
The shape of $p_1^M(\theta|b,h)$ can be described as follows:
it is zero at the null value $\theta_0=0.25$, as required,
it increases rapidly as $\theta$ goes to 1, while it goes up more gently
as $\theta$ goes to zero.
It is clear that this moment prior will not be suitable for testing purposes,
because it puts too much mass away from $\theta_0$. This is confirmed
by the
thin dashed line in Figure~\ref{figbernoullifirst}(b): the null model
is unduly favored. The thin dotted line in the same plot shows that
things get even worse for $h=2$ and $t=0$; again,
for now, please disregard the curves with $t=8$ and $t=13$.
On the other hand, the moment prior has an improved learning rate as
the sample size grows:
we postpone the illustration of this feature to the next section,
after the moment prior has been made suitable for testing purposes
by means of a probability mass transfer toward the null value $\theta_0$.
\end{example}
%

Rousseau and Robert, in discussing \citet{ConsLaro2011},
raise an interesting point in relation to moment priors.
They cast the problem in a decision-theoretic setup and use the
well-known duality
between prior and loss function (\cite{Rubi1987}; \cite{Robe2001})
to suggest that nonlocal priors should be replaced by the use of
suitable loss functions,
which take into account the distance from the null.
This perspective was actually pursued in \citet{RobeCase1994};
see also \citet{GoutRobe1998}.
Indeed, it can be checked that the optimal Bayesian decision,
under a $\{0,1\}$-loss function and a moment prior of the form
$p^M_1(\xi_1) \propto(\xi_1-\xi_0)^{2h}p_1(\xi_1)$,
where $p_1(\xi_1)$ is a local prior and $\xi_0$ a null parameter value,
coincides with that arising from the local prior $p_1(\xi_1)$
and a ``distance weighted'' loss function of the form
\[
L(a,\xi)= \cases{ K_1, & if $a=1$ and $\xi=
\xi_0$,
\cr
(\xi-\xi_0)^{2h}, & if $a=0$ and $
\xi\neq\xi_0$,
\cr
0, & otherwise,}
\]
where $a$ is the action, taking value 0 or 1 if the chosen model is
${\cal M }_0$ or ${\cal M }_1$, respectively,
while $ K_1 = \vat_1[(\xi_1-\xi_0)^{2h}]$ is the expected
loss, under the local prior $p_1(\xi_1)$, when $\mathcal{M}_0$ is
wrongly chosen.
This interpretation of moment priors is interesting and, from our viewpoint,
it reinforces their usefulness, because it shows that a moment prior
can be
justified using decision theory.
%
%

\section{Intrinsic Moment Priors}
\label{secintrimom}

Example~\ref{exbernoullictd} shows that the moment prior obtained
from a default local prior, call it the \emph{default moment prior},
does not accumulate enough mass around the null value
(more generally around the subspace specified by the null model).
This suggests applying the intrinsic procedure to the default moment prior,
obtaining in this way a new class of priors for testing nested hypotheses,
which we name intrinsic moment priors.
The improved learning rate extends to the latter priors, because each
of them is a mixture (through the intrinsic procedure) of nonlocal priors.

Our strategy for a balanced objective Bayesian comparison of two nested models
thus starts with a default prior under each of the two models and then envisages
two steps:
(i) construct the default moment prior of order $h$ under the larger model;
%
(ii) for a given training sample size $t$,
generate the corresponding intrinsic moment prior.
We recommend using the resulting prior to compute the BF:
step (i) improves the learning rate (when the null model holds),
while step (ii) makes sure that the testing procedure exhibits
a good small sample behavior in terms of
the evidence curve.
%
%
%
We first illustrate intrinsic moment priors in our running example,
then we discuss the choice of $t$. In Section~\ref{sectwoBinomial}
the procedure will be implemented to test the equality of two proportions,
while in Section~\ref{seclogreg} it will be developed for
the family of logistic regression models.

\begin{example}[(Bernoulli c.t.d.)]
Recall that the intrinsic prior is an average of ``posterior'' distributions.
Since in our case we start from the moment prior (\ref{MC})
with default choice $a_1=a_2=b$, the intrinsic moment prior for $\theta$
with training sample size $t$ will be given by
%
\begin{eqnarray}
\label{pIM}
&&p_1^{\mathrm{IM}}(\theta|b,h,t)\nonumber\\
&&\quad=\sum
_{x=0}^t \frac{(\theta-\theta_0)^{2h}\operatorname{Beta}(\theta
|b+x,b+t-x)}{K(b+x,b+t-x,h,\theta_0)} \\
&&\hspace*{15.6pt}\qquad{}\cdot\operatorname{Bin}(x|t,
\theta_0),\nonumber
\end{eqnarray}
where $K(\cdot,\cdot,h,\theta_0)$ is defined in (\ref{Kgeneral}),
and we exploited conjugacy.
Notice that (\ref{pIM}) describes a family of prior distributions
including the standard intrinsic prior \mbox{($h=0$)}
and the default prior ($h=0,t=0$) as special cases.
Similarly, from (\ref{eqintriGeneralBF}) we find
%
\begin{eqnarray}
\label{eqintriMomBerBF}
&&\mathrm{BF}_{10}^{\mathrm{IM}}(y|b,h,t)\nonumber\\
&&\quad=\sum
_{x=0}^t 
\mathrm{BF}_{10}^M(y|b+x,b+t-x,h)\\
&&\hspace*{15.6pt}\qquad{}\cdot
\operatorname{Bin}(x|t,\theta_0),\nonumber
\end{eqnarray}
where $\mathrm{BF}_{10}^M(y|\cdot,\cdot,h)$ is defined in (\ref{eqmomentBFbern}).\vadjust{\goodbreak}

Figure~\ref{figbernoullifirst}(a) shows (letting $b=1$)
the effect of applying the intrinsic procedure
to the default moment prior of order $h=1$ (dashed curves):
as $t$ grows, the overall shape of the prior density changes considerably,
because more and more probability mass in the extremes
is displaced toward $\theta_0$, giving rise to two modes,
while the nonlocal nature of the prior is preserved, because the
density remains
zero at $\theta_0=0.25$. In this way, as shown in Figure \ref
{figbernoullifirst}(b),
the evidence against the null for small samples is brought back to
more reasonable values (with respect to the default moment prior).
More specifically, Figure~\ref{figbernoullifirst}(b) shows that
the intrinsic moment prior with $h=1$ and $t=8$ (a choice explained
later in Section~\ref{subsecChoosingTrainingSampleSize})
performs comparably to the uniform prior (and to the standard intrinsic prior
with unit training sample) over a broad range of values for
the observed sampling fraction $\bar{y}$;
this intrinsic moment prior results in a smaller amount of evidence
(against~${\cal M}_0)$
for values of $\bar{y}$ close to $0.25=\theta_0$, which is to be
expected for continuity,
but induces a steeper evidence gradient as $\bar{y}$ moves away
from the null point in either direction, which makes it appealing.

\begin{figure*}

\includegraphics{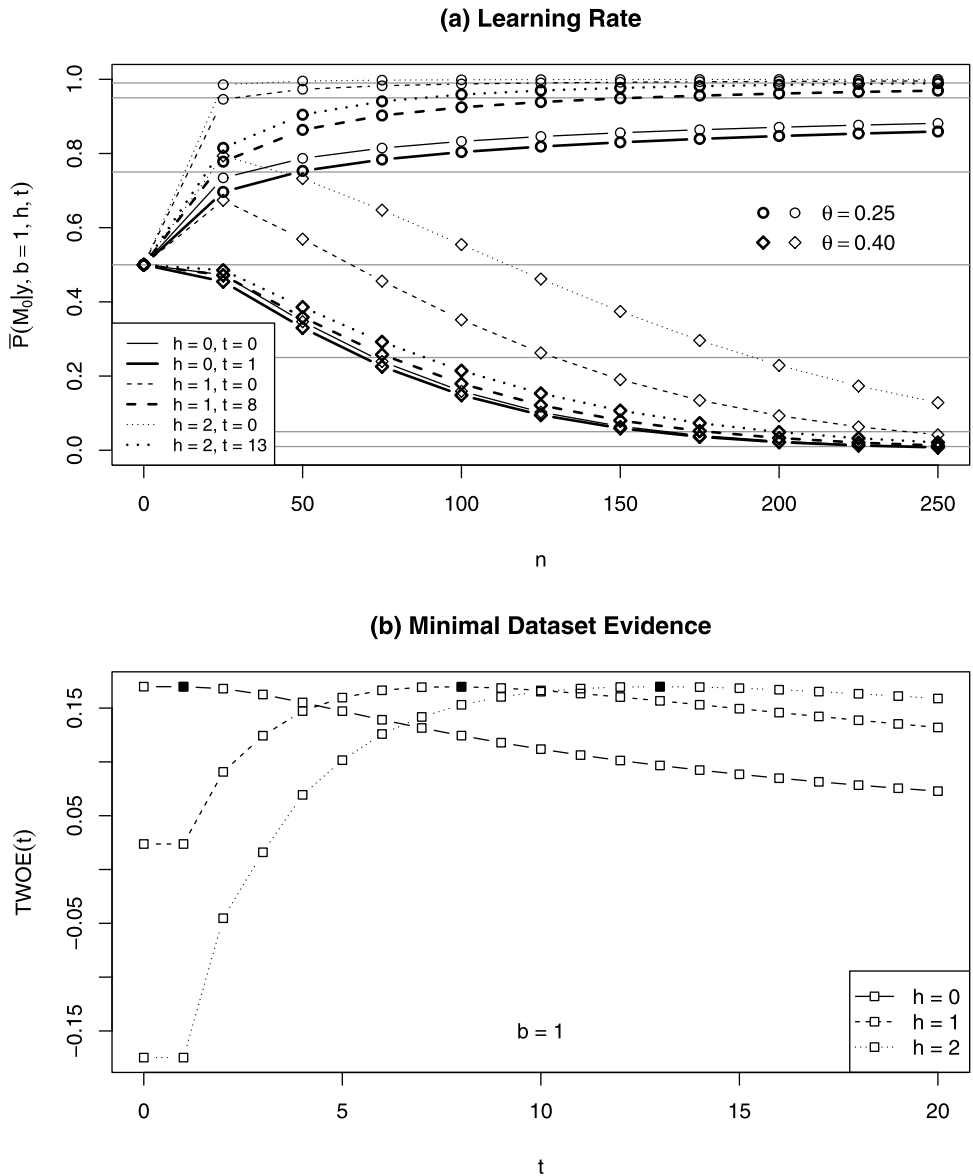}

\caption{Learning rate \textup{(a)} and minimal data set evidence
\textup{(b)}
for the Bernoulli example. Horizontal gray lines in \textup{(a)} denote
possible decision thresholds at 1\%, 5\%, 25\%, 50\%, 75\%, 95\% and
99\%
on the posterior probability scale.}\vspace*{6pt}
\label{figbernoullisecond}
\end{figure*}

The learning rate of the intrinsic moment prior
is illustrated in Figure~\ref{figbernoullisecond}(a),
which reports the average posterior probability
of the null model when $\theta=0.25$ (null value)
and when $\theta=0.4$ (an instance of the alternative model).
It is apparent from this plot that
a nonlocal prior ($h>0$) is needed, if strong evidence in favor of the null
has ``ever'' to be achieved, and also that the intrinsic procedure is crucial
to calibrate small sample evidence.
These results are striking, and they signal that our strategy actually
represents a marked improvement over current methods.
Notice that there is an associated cost:
the moment prior trades off a delay in learning the alternative model
for speed in learning the null model; the intrinsic procedure
is remarkably effective in controlling this trade-off.
In light of Figure~\ref{figbernoullisecond}(a),
we recommend letting $h=1$ by default and trying $h=2$ for sensitivity purposes;
we remark that $h=1$ is enough to change the convergence rate of
$\mathrm{BF}_{10}^{\mathrm{IM}}(y^{(n)})$,
when ${\cal M}_0$ holds, from sub-linear to super-linear.
\end{example}



\subsection{Choosing the Training Sample Size}
\label{subsecChoosingTrainingSampleSize}


Recall that the goal of the intrinsic procedure is to transfer
probability mass
toward the null subspace within the default prior under ${\cal M}_1$.
There is clearly
a tension here between this aim and that of leaving enough mass in
other areas
of the parameter space, not to unduly discredit ${\cal M}_1$.
This is precisely the issue we face when choosing $t$.
We now\vadjust{\goodbreak} provide some guidelines for the Bernoulli problem,
with a view to more general situations.


We aim at a single recommended value of $t$ for all possible values of
$\theta_0$.
For this purpose, we fix $\theta_0=1/2$, representing the worst case
scenario in terms of
the information content of a single observation.
A minimal sample size to discriminate between $\mathcal{M}_1$ and
$\mathcal{M}_0$ is $n=2$,
with possible data values $y=0, 1, 2$.
Define the \emph{weight of evidence} against the null using an
intrinsic moment prior
as $\mathrm{WOE}_y(t)=\log \mathrm{BF}^{\mathrm{IM}}_{10}(y|b,h,t)$,
where we focus on the dependence on $t$
for a given choice of $b$ and $h$.
By symmetry, $\mathrm{WOE}_0(t)=\mathrm{WOE}_2(t)$.
However, $\mathrm{WOE}_1(t) \neq\break \mathrm{WOE}_0(t)$;
this is why we are able to discriminate between the two models if $n=2$,
which would not happen with $n=1$.
It can be checked that $\mathrm{WOE}_1(t)$ is increasing in $t$,
while $\mathrm{WOE}_0(t)=\mathrm{WOE}_2(t)$ is decreasing in $t$.
The explanation of this phenomenon is simple, keeping in mind that
an increase in $t$ transfers probability mass toward $\theta_0=1/2$
within the prior:
the value $y=1$ supports ${\cal M}_0$, and thus its marginal probability
under ${\cal M}_1$ will increase with $t$;
the values $y=0$ and $y=2$ support ${\cal M}_1$, and thus an increase
in $t$
will make their marginal probability under ${\cal M}_1$ smaller.
%
%
%
How far should we let $t$ grow?
%
%
To answer this question, define the \emph{total weight of evidence}
$\operatorname{TWOE}(t)=\sum_y\mathrm{WOE}_y(t)$, and consider the weight of evidence as a
sort of currency:
we will be willing to trade off a decrease in $\mathrm{WOE}_0(t)$ and $\mathrm{WOE}_2(t)$
for an increase in $\mathrm{WOE}_1(t)$ as long as we get more than we give,
that is, as long as we increase $\operatorname{TWOE}(t)$.
Define $t^*=\arg\max_t \operatorname{TWOE}(t)$ and assume that this quantity is
well-defined.
The value $t^*$ represents our optimal training sample size
when implementing the intrinsic procedure.
Since this choice of $t$ is based on a
somewhat unusual criterion,
we will be willing to let $t$ vary in a neighborhood of $t^*$
for a sensitivity analysis.

We remark that the above strategy to find $t^*$ in an intrinsic procedure
is general, at least for finitely discrete data models.
In particular, it can be used to determine an optimal training
sample size also for the standard intrinsic prior ($h=0$).
Figure~\ref{figbernoullisecond}(b) plots $\operatorname{TWOE}(t)$ for $h=0,1,2$,
assuming a uniform default prior ($b=1$).
Interestingly, when $h=0$ (standard intrinsic prior),
we find $t^* \in\{0,1\}$.
This seeming indeterminacy can be explained by noticing that,
when $\theta_0=0.5$, the intrinsic prior with $t=1$ is the uniform prior,
that is, it is the same as the default prior (corresponding to $t=0$).
On the other hand, when the starting prior is the default moment prior
of order $h=1$,
it turns out that $t^*=8$, while for $h=2$ we obtain $t^*=13$,
so that with nonlocal moment priors the intrinsic procedure is necessary:
this makes sense, because the starting prior puts mass at the endpoints
of the parameter space in a rather extreme way.

\section{Testing the Equality of Two Proportions}
\label{sectwoBinomial}

Suppose the larger (encompassing) model is the product of two binomial models
%
\begin{eqnarray}\label{eqlikelihood}
&&\mathcal{M}_1\dvtx  f_1(y_1,y_2|
\theta_1,\theta_2)\nonumber\\[-8pt]\\[-8pt]
&&\quad= \operatorname{Bin}(y_1|n_1,
\theta_1) \operatorname{Bin}(y_2|n_2,
\theta_2),\nonumber 
\end{eqnarray}
where $n_1$ and $n_2$ are fixed sample sizes.
The null model assumes $\theta_1=\theta_2=\theta$, so that
%
\begin{equation}
\mathcal{M}_0\dvtx  f_0(y_1,y_2|
\theta) = \operatorname{Bin}(y_1|n_1,\theta)
\operatorname{Bin}(y_2|n_2,\theta).\hspace*{-25pt}
\end{equation}
%
A default prior for $\theta$ under $\mathcal{M}_0$ is
$p_0(\theta|b_0)=\operatorname{Beta}(\theta|\break b_0,b_0)$,
while a default prior for $(\theta_1,\theta_2)$ under $\mathcal{M}_1$
is given by $p_1(\theta_1,\theta_2|b_1,b_2)=\operatorname{Beta}(\theta_1|b_1,b_1)
\operatorname{Beta}(\theta_2|\break b_2,b_2)$.

Starting from a more general conjugate prior\break
$\operatorname{Beta}(\theta_1|a_{11},a_{12})\operatorname{Beta}(\theta_2|a_{21},a_{22})$
under $\mathcal{M}_1$, we define the moment prior of order $h$ as
%
\begin{eqnarray}\label{eqpriormoment}
&&
p_1^M(\theta_1,\theta_2|a,
h)\nonumber\\
&&\quad= \frac{(\theta_1-\theta_2)^{2h}}{K(a,h)}
\operatorname{Beta}(\theta_1|a_{11},a_{12})\\
&&\qquad{}\cdot
\operatorname{Beta}(\theta_2|a_{21},a_{22}),\nonumber
\end{eqnarray}
where $a = [[a_{jk}]_{k=1,2}]_{j=1,2}$ is a matrix of strictly positive
real numbers and
%
\begin{eqnarray}\label{eqnormaconst}
K(a,h)&=&\sum_{j=0}^{2h}\pmatrix{2h
\cr
j}(-1)^j \frac{B(a_{11}+j,a_{12})}{B(a_{11},a_{12})}
\nonumber\\[-8pt]\\[-8pt]
&&\hspace*{15.4pt}{}\cdot\frac{B(a_{21}+2h-j,a_{22})}{B(a_{21},a_{22})}.\nonumber
\end{eqnarray}
%
The default moment prior will be obtained by letting $a_{11}=a_{12}=b_1$
and $a_{21}=a_{22}=b_2$; letting $h=0$ will then return, as usual, the
default prior.

Consider now the intrinsic approach applied to the default moment prior.
Since the data consist of two counts, a vector of length two is needed to
specify the training sample size. The \emph{intrinsic moment} prior of
order $h$
with training sample size $t=(t_1,t_2)$ will be defined as
%
\begin{eqnarray}\label{eqpriorintrinsic}\quad
&&
p_1^{\mathrm{IM}}(\theta_1,\theta_2|b,
h, t)\nonumber\\[-8pt]\\[-8pt]
&&\quad= \sum_{x_1=0}^{t_1}\sum
_{x_2=0}^{t_2} p_1^{M}\bigl(
\theta_1,\theta _2|a_x^\star, h
\bigr) m_0(x_1,x_2|b_0),\nonumber
\end{eqnarray}
where $b=(b_0,b_1,b_2)$,
while $(a_x^\star)_{11}^{}=b_1+x_1$, $(a_x^\star)_{12}^{}=b_1+t_1-x_1$,
$(a_x^\star)_{21}^{}=b_2+x_2$, $(a_x^\star)_{22}^{}=b_2+t_2-x_2$,
and
%
\begin{eqnarray}\label{eqmarliknested}\quad
&&
m_0(x_1,x_2|b_0)\nonumber\\
&&\quad=
\pmatrix{t_1
\cr
x_1}\pmatrix{t_2
\cr
x_2}\\
&&\qquad{}\cdot\frac
{B(b_0+x_1+x_2,b_0+t_1+t_2-x_1-x_2)}{B(b_0,b_0)};\nonumber
\end{eqnarray}
letting $h=0$ returns the standard intrinsic prior
$p_1^{I}(\theta_1,\theta_2|b,t)$.

The BF against $\mathcal{M}_0$ using the intrinsic moment prior
under $\mathcal{M}_1$ is given by
%
\begin{eqnarray}\label{eqbfintrinsic}\qquad
&&
\mathrm{BF}_{10}^{\mathrm{IM}}(y_1,y_2|b, h, t)\nonumber\\[-8pt]\\[-8pt]
&&\quad=
\sum_{x_1=0}^{t_1}\sum
_{x_2=0}^{t_2} \mathrm{BF}_{10}^{M}
\bigl(y_1,y_2|a^\star_x, h\bigr)
m_0(x_1,x_2|b_0),\nonumber
\end{eqnarray}
where $\mathrm{BF}_{10}^{M}(y_1,y_2|a^\star_x, h)$ is the BF obtained with the
``posterior''
$p_1^{M}(\theta_1,\theta_2|a_x^\star, h)$ as parameter prior under
$\mathcal{M}_1$
(and the default parameter prior under~$\mathcal{M}_0$).

Similarly to the Bernoulli case, we can write
%
\begin{equation}\label{eqbfmoment}\qquad
\mathrm{BF}_{10}^{M}(y_1,y_2|a, h)=
\frac{K(a_y^\star,h)}{K(a,h)}\mathrm{BF}_{10}(y_1,y_2|a),
\end{equation}
where $(a_y^\star)_{11}^{}=a_{11}+y_1$,
$(a_y^\star)_{12}^{}=a_{12}+n_1-y_1$,\break
$(a_y^\star)_{21}^{}=a_{21}+y_2$, and $(a_y^\star)_{22}^{}=a_{22}+n_2-y_2$.
A standard computation then gives
\begin{eqnarray*}
&&
m_1(y_1,y_2|a)\\
&&\quad=
\pmatrix{n_1
\cr
y_1}\pmatrix{n_2
\cr
y_2} \\
&&\qquad{}\cdot B(a_{11}+y_1, a_{12}+n_1-y_1)\\
&&\qquad{}\cdot B(a_{21}+y_2, a_{22}+n_2-y_2)\\
&&\qquad{}/\bigl(B(a_{11}, a_{12})B(a_{21}, a_{22})\bigr), 
\end{eqnarray*}
and it follows that the Bayes factor against $\mathcal{M}_0$
obtained with the moment prior under $\mathcal{M}_1$
(and the default prior under $\mathcal{M}_0$) can be written as
\begin{eqnarray*}
&&
\mathrm{BF}_{10}(y_1,y_2|a)\\
&&\quad=
B(b_0,b_0)B(a_{11}+y_1, a_{12}+n_1-y_1)\\
&&\qquad{}\cdot B(a_{21}+y_2, a_{22}+n_2-y_2)\\
&&\qquad{}/
\bigl(B(a_{11}, a_{12})B(a_{21}, a_{22})\\
&&\hspace*{9pt}\qquad{}\cdot B(b_0+y_1+y_2,b_0+n_1+n_2-y_1-y_2)\bigr). 
\end{eqnarray*}
Using the above expression
in (\ref{eqbfmoment}) and plugging
the latter into (\ref{eqbfintrinsic}) provides an explicit expression for
$\mathrm{BF}_{10}^{\mathrm{IM}}(y_1,y_2|b, h, t)$.

\subsection{Choice of Hyperparameters}
\label{choiceofhypars}

The intrinsic moment prior $p_1^{\mathrm{IM}}(\theta_1,\theta_2|b, h, t)$
depends on three hyperparameters.
We recommend choosing $b_1+b_2=b_0$,
so that the same amount of prior information is imposed under $\mathcal{M}_1$,
on the vector parameter $(\theta_1,\theta_2)$,
and under $\mathcal{M}_0$, on the scalar parameter $\theta$.
Specifically, adopting a prior distribution with unit prior information,
we let $b_0=1/2$, and $b_1=b_2=1/4$,
for the balanced case $n_1=n_2$, while in the nonbalanced case
$b_1$ and $b_2$ will be proportional to $n_1$ and $n_2$.
Then, as in the Bernoulli example, we recommend choosing
$h=1$, which is enough to change the asymptotic learning rate
of the BF, when the null holds, from sub-linear to super-linear.
%
Finally, concerning the choice of $t$, we follow the general procedure
outlined in the Bernoulli example, with suitable specific modifications
to deal with the present case. In particular, we focus on the balanced case\vadjust{\goodbreak}
to obtain a single optimal value of $t_+=t_1+t_2$, which can then be used
also in the nonbalanced case to specify $t_1$ and $t_2$ as (approximately)
proportional to $n_1$ and $n_2$.

Clearly, $n_1=n_2=1$ represent the minimal sample sizes for the testing
problem at hand.
In this case, of the four possible data outcomes, two are supportive
for ${\cal M}_0$, namely, $(y_1=0,y_2=0)$ and $(y_1=1,y_2=1)$,
and two are supportive for ${\cal M}_1$, namely, $(y_1=0,\break y_2=1)$ and
$(y_1=1,y_2=0)$.
We repeat the argument in Section~\ref{subsecChoosingTrainingSampleSize}
and take $t_+^*=\arg\max_{t_+} \operatorname{TWOE}(t_+)$ as the optimal total
training sample size,
where\break $\operatorname{TWOE}(t_+)=\sum_y\mathrm{WOE}_y(t_+)$ and
$\mathrm{WOE}_y(t_+) =\break \log{\mathrm{BF}_{10}^{\mathrm{IM}}(y_1,y_2|b, h, t)}$
with $t=(t_+/2, t_+/2)$ and $t_+$ even.

Figure~\ref{figCharact}(a) plots $\operatorname{TWOE}(t_+)$ for $h=0,1,2$.
As in the Bernoulli case, $t_+^*$ is well defined and when $h=0$
(standard intrinsic prior) we get $t_+^*=0$.
Hence, in this case, we would recommend a sensitivity analysis
in line with that carried out by Casella and Moreno
(\citeyear{CaseMore2009}, Table 2).
%
%
On the other hand, when
$h=1$ we find $t_+^*=8$, while for $h=2$ we get $t_+^*=14$;
as in the Bernoulli case, it turns out that,
starting with a nonlocal moment prior, the intrinsic approach is needed.
In the following subsection we highlight some features of the intrinsic
moment priors
specified by the above values of $h$ and $t_+=t_+^*$ (including $h=0$
and $t_+^*=0$).

\subsection{Characteristics of Intrinsic Moment Priors}
\label{subseccharimps}

\begin{figure*}[t]

\includegraphics{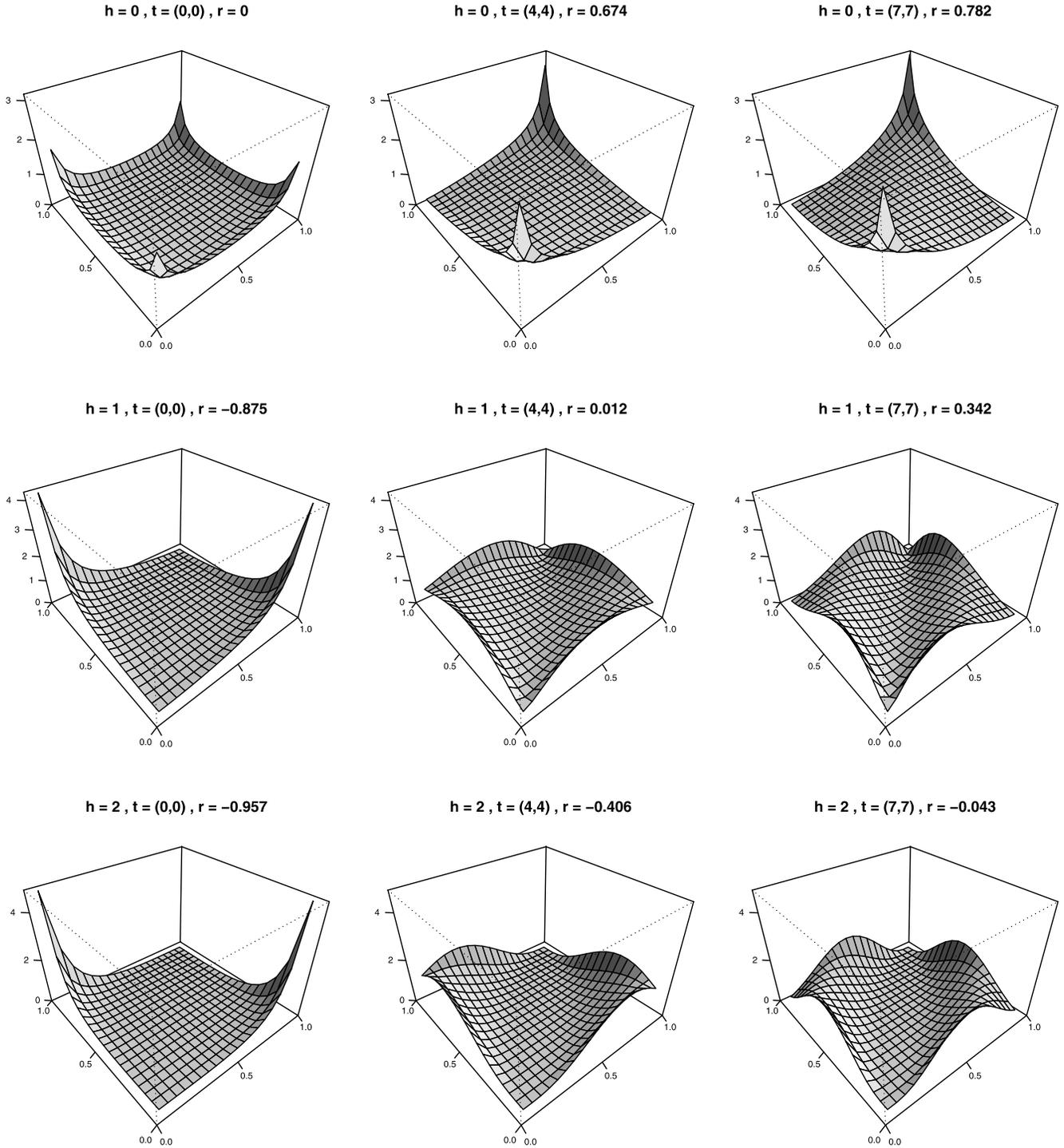}

\caption{Intrinsic moment prior densities for comparing two proportions
($b_0=1/2$, $b_1=1/4$).}\vspace*{-2pt}
\label{figPriorsRev}
\end{figure*}

Figure~\ref{figPriorsRev} presents a collection of nine priors
for $(\theta_1,\theta_2)$ under ${\cal M}_1$, each labeled with
its corresponding correlation coefficient $r$. Although the absolute
values of $r$ are of dubious utility in describing these distributions,
because of their shape, comparison of the displayed values enables us
to highlight the roles played by $h$ and $t_+$: as $h$ grows
the prior mass is displaced from areas around the line $\theta
_1=\theta_2$
to the corners ($\theta_1=0,\theta_2=1$) and ($\theta_1=1,\theta_2=0$),
thus inducing negative correlation; on the other hand,
as $t_+$ grows the prior mass is pulled back toward either side
of the line $\theta_1=\theta_2$, and positive correlation is induced.
The priors in the first row are local,
while those in the second and third rows are nonlocal.
The three distributions on the main diagonal represent,
for the three values of~$h$, our suggested priors
based on the criterion for the choice of $t_+$
described in Section~\ref{choiceofhypars}.
Notice that $r\simeq0$ for all three suggested priors,
so that the chosen value of $t_+$ can be seen as ``compensating'' for $h$.
%

\begin{figure*}

\includegraphics{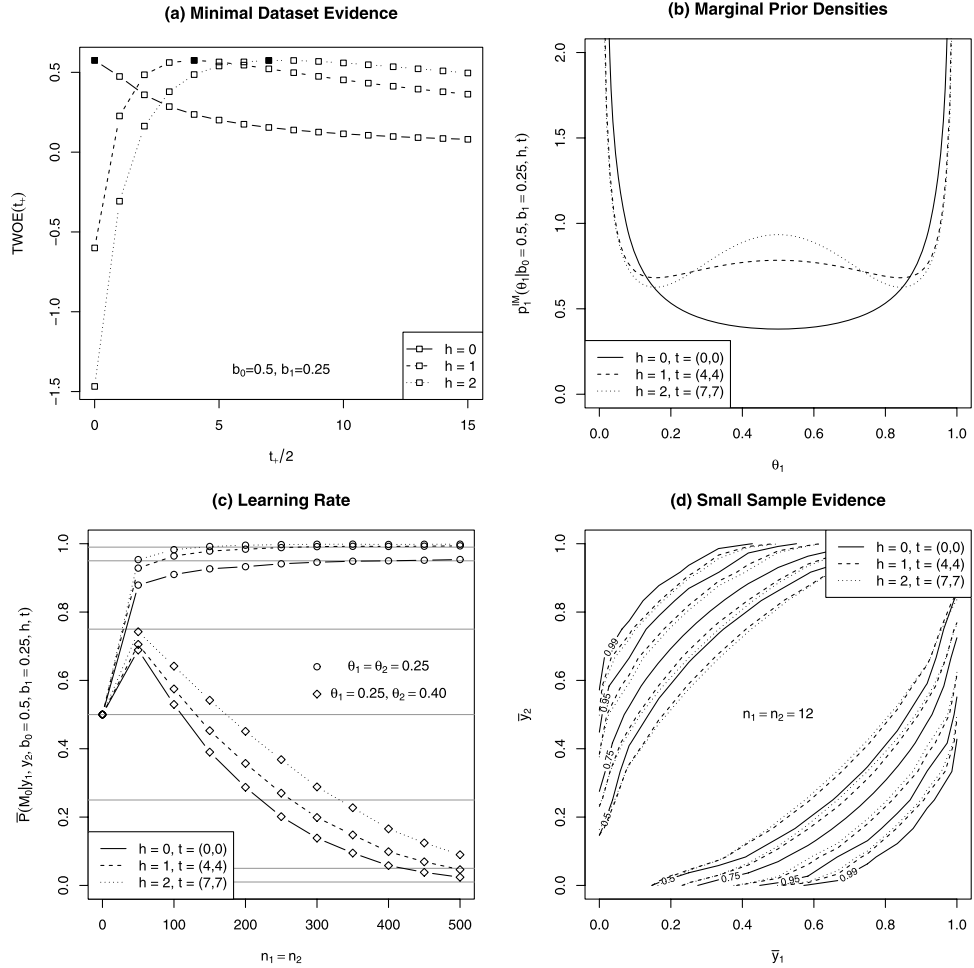}

\caption{Characteristics of intrinsic moment priors for comparing two
proportions.
Horizontal gray lines in \textup{(c)} denote possible decision thresholds
at 1\%, 5\%, 25\%, 50\%, 75\%, 95\% and 99\% on the posterior
probability scale.
Contour lines in \textup{(d)} refer to the posterior probability of the
alternative model
computed from data $y_1=n_1\bar{y}_1$ and $y_2=n_2\bar{y}_2$
(letting $b_0=1/2$ and $b_1=1/4$).}
\label{figCharact}
\end{figure*}

Some further insight into the structure of the priors
on the main diagonal of Figure~\ref{figPriorsRev}
can be gleaned by looking at Figure~\ref{figCharact}(b),\vadjust{\goodbreak}
which reports their mar\-ginal distributions
(identical for $\theta_1$ and $\theta_2$).
All three densities are symmetric around the value $0.5$,
but the two intrinsic moment priors with $h>0$
give more credit to the inner values of the interval $(0,1)$.
%
For these three priors, Figure~\ref{figCharact}(c) reports
the average posterior probability of the null model computed on 1000
simulated data sets
of increasing size, generated first letting $\theta_1=\theta_2=0.25$
and then setting $\theta_1=0.25$, $\theta_2=0.4$ (an instance of the
alternative model).
Notice that, while in the Bernoulli example we were able to implement
an exact computation,
in this case we had to resort to a Monte Carlo approximation, because
exact computation
would have been too demanding (at least for ordinary computational resources).

The learning rate is quite different under the three priors
on the main diagonal of Figure~\ref{figPriorsRev}. Like in the
Bernoulli example,
when the data are generated under the null model
a much quicker correct response is provided by the nonlocal priors:
for sample sizes up to 500
the average posterior probability of ${\cal M}_0$
under the default prior hardly reaches the 95\% threshold, whereas under
the nonlocal intrinsic moment priors it easily achieves the 99\% threshold
by the time 250 observations have been collected.
On the other hand, switching from $h=0$ to $h>0$,
the learning rate under the alternative model
is compromised in the short run, but not in the long run.
%
%

Figure~\ref{figCharact}(d) illustrates the small sample behavior
of intrinsic moment priors, by reporting the contour lines in the
$(\bar{y}_1,\bar{y}_2)$-plane of observed frequencies,
when $n_1=n_2=12$, for selected thresholds of the posterior probability
of ${\cal M}_1$.
There is a clear indication that the displayed thresholds
are reached for pairs $(\bar{y}_1,\bar{y}_2)$
closer to the $\bar{y}_1=\bar{y}_2$ line under the nonlocal priors
than under the default prior. Similarly to the Bernoulli example,
this is due to the steeper gradient of the evidence surface as the data
move away from the null supporting values.

\section{Variable Selection in Logistic Regression Models}
\label{seclogreg}

We now develop the intrinsic moment procedure when the models under
comparison are
logistic regression models. This demonstrates that the general
procedure can be applied
to a flexible and general class of discrete data models.

Suppose we observe $N$ independent binomial observations,
$y=(y_1,\ldots, y_N)$,
where
\[
y_i \g\theta_i \sim\operatorname{Bin}(n_i,
\theta_i);\quad i=1,\ldots, N. %
\]
The binomial probabilities $\theta=(\theta_1,\ldots, \theta_N)$
are assumed to depend on the values of $k$ explanatory variables
$z_{ij}$, $i=1,\ldots, N$, $j=1,\ldots, k$,
through linear predictors $\eta=(\eta_1,\ldots, \eta_N)$,
where
\[
\eta_i= \log\frac{\theta_i}{1-\theta_i}=\beta_0+\sum
_{j=1}^k z_{ij}\beta _j;\quad i=1,\ldots, N. %
\]
Hence, the likelihood is
$f(y \g\beta)= \{ \prod_{i=1}^N {n_i\choose y_i}  \}
L(\beta\g\break y, n)$,
where
\begin{eqnarray*}
&&
L(\beta\g y,n) \\
&&\quad= \prod_{i=1}^N\exp \Biggl\{ y_i \Biggl(\beta_0 +
\sum_{j=1}^k z_{ij}
\beta_j \Biggr)\\
&&\hspace*{35.6pt}\qquad{}-n_i \log \Biggl(1+\exp \Biggl[
\beta_0+\sum_{j=1}^k
z_{ij}\beta_j \Biggr] \Biggr) \Biggr\}, %
\end{eqnarray*}
$\beta=(\beta_0, \beta_1,\ldots, \beta_k)$ and $n=(n_1,\ldots, n_N)$.
We refer to this model as the full model.
Further models under consideration for variable selection
correspond to an exclusion of some explanatory variables,
that is, to setting some $\beta_{j}=0$ ($j\neq0$).

In the development below, we present the prior for the full model
with $k$ explanatory variables, but the prior for any other model takes
an identical form with a regression parameter of correspondingly lower
dimensionality.

For convenience, and consistency with our earlier developments
in the context of two binomial models, we adopt a conjugate local prior
(\cite{BedrEtAl1996}) given by $p^C(\beta\g u,w)\propto L(\beta\g u,w)$,
where $u=(u_1,\ldots, u_N)$ and $w=(w_1,\ldots,\break w_N)$ are
hyperparameters corresponding
to $y=(y_1,\break\ldots, y_N)$ and $n=(n_1,\ldots, n_N)$, respectively,
in the likelihood.
Letting $w_+=\sum_{i=1}^N w_i$,
we choose as the default prior specification
%
\begin{equation}
\label{eqwiAndui} w_i=w_+ \frac{n_i}{\sum_i n_i },\quad u_i=
\frac{w_i}{2},
\end{equation}
where $w_+$ represents a prior sample size.
The condition $u_i=w_i/2$ ensures that the mode of the prior is at
$\beta=0$.
To see why, recall that the prior, as a function of $\beta$,
is proportional to the likelihood.
Now, if $y_i=n_i/2$, then the MLE of each $\theta_i$, unconstrained by
the model,
is exactly $1/2$ and, therefore, the MLE of each $\eta_i$ is zero.
The value $\eta=0$ is attained within any logistic regression model by
$\beta=0$
and, hence, this value must also maximize the model constrained likelihood,
which corresponds to the prior density. As a default choice,
corresponding to unit prior information, we take $w_+=1$.
%
For the comparison of two proportions $\theta_1$ and $\theta_2$
($N=2$) in the balanced case $n_1=n_2$,
this formulation leads to identical default local priors
$\theta_i \sim \operatorname{Beta}(1/4,1/4)$ with $\theta_1$ and $\theta_2$ independent.

%

In order to construct the moment prior,
we need to specify a function $g_h(\beta)$.
We choose
%
\begin{equation}
\label{eqgOfbeta} g_h(\beta)= \prod_{j=1}^k
\beta_j^{2h},
\end{equation}
which vanishes if at least one $\beta_j=0$ ($j\neq0$),
implying that we separate the full model from every model nested within it
having one less explanatory variable.
In the context of variable selection for Gaussian distributions,
this choice of $g_h(\beta)$ has been used by \citet{ConsLaro2011}
and also by \citet{JohnRoss2012}, who named the resulting
nonlocal prior
a \emph{product moment} prior.
Our main result in the \hyperref[app]{Appendix} (Theorem~\ref{thmmain}),
though stated for i.i.d. observations,
confirms that this is a sensible choice for variable selection,
resulting in an effective separation of models.
With this choice we obtain\looseness=1
\[
p^M(\beta\g u,v,h) \propto p^C(\beta\g u,w) \prod
_{j=1}^k \beta_j^{2h}.
%
\]\looseness=0

At this stage, to specify the intrinsic moment prior under any given
model $\mathcal{M}$,
we need a reference mod\-el~$\mathcal{M}_0$,
which we take as the null model having no explanatory variable ($k=0$)
because it is nested in every other model.
In this construction, the priors used in any pairwise model comparison
depend only on the (common) null model.
This strategy is called \emph{encompassing from below}
and provides a coherent model comparison procedure; see \citet
{LianPauletal2008}.
Under $\mathcal{M}_0$ we assume a default prior for the intercept
$\beta_0$
given by
\[
p(\beta_0) \propto\exp\bigl\{ \beta_0 u_+ -w_+ \log
\bigl(1+\exp[\beta_0]\bigr) \bigr\}, %
\]
where $u_+=\sum_{i=1}^N u_i$,
which corresponds to a $\operatorname{Beta}(u_+,\break w_+-u_+)=\operatorname{Beta}(1/2,1/2)$ distribution,
because of the assumed value $w_+=1$, for the common success probability
implied by $\mathcal{M}_0$ in the comparison of two proportions.

The final step in the construction of the intrinsic moment prior requires
the specification of training samples,
which also involves covariates when dealing with regression models.
Methods for choosing covariates for training data have been discussed
for Gaussian regression models by \citet{GiroEtAl2006}.
We assume that the covariate patterns in the training data are a subset
of those appearing in the observed data.
Following formula (\ref{intriGeneral}),
we now construct the intrinsic moment prior for the parameter of a
logistic regression model.
Let
\begin{eqnarray*}
&&
p^M(\beta\g x+u,t+w,h) \\
&&\quad\propto 
\Biggl\{\prod
_{j=1}^k \beta_j^{2h} \Biggr
\}L(\beta\g x+u,t+w) %
\end{eqnarray*}
be the posterior moment prior based on
the training sample $x=(x_1,\ldots,x_N)$, with
training sample sizes given by $t=(t_1,\ldots,t_N)$,
where some of the $t_i$ may be zero.
Since $x$ is drawn from $m_0(x)$,
the marginal joint distribution under $\mathcal{M}_0$,
the intrinsic moment prior is thus given by
\begin{eqnarray*}
&&
p^{\mathrm{IM}}(\beta\g h, t)\\
&&\quad=\sum_x
m_0(x) \frac
{\{\prod_{j=1}^k \beta_j^{2h} \}
L(\beta\g x+u,t+w)}{Q(x+u,t+w, h)},
\end{eqnarray*}
where\vspace*{2pt} $Q(z,s, h)= \int_{\Re^{k+1}} \{\prod_{j=1}^k \beta_j^{2h} \}
L(\beta\g z,s) \,d\beta$.
The existence of $Q(z,s, h)$ follows from the theorem in
\citeauthor{Fors2010} (\citeyear{Fors2010}, Section 6)
stating that a necessary and sufficient condition for a log-concave function
over ${\cal R}^d$ to have a finite integral is that it achieves its maximum
in the interior of the parameter space. Here, we need to adapt this
result slightly.
First, we notice that in each (open) orthant
the integrand is log-concave, because both its constituent components
are log-concave:
$\prod_{j=1}^k \beta_j^{2h}$ by straightforward calculus,
and $L(\beta\g x+u,t+w)$ by log-concavity of the likelihood
for a binomial logistic regression model.
For our default choices of $u$ and $w$
(or any alternative choice with $u>0$ and $0<u<w$),
$L(\beta\g x+u,t+w)$ has a unique finite maximum,
provided that the model is identified (which we will assume).
Hence, $L(\beta\g x+u,t+w)$ tends to zero, as $\|\beta\|\to\infty$,
in any direction, and so does $\{\prod_{j=1}^k \beta_j^{2h} \}L(\beta
\g x+u,t+w)$,
due to the dominance of $L(\beta\g x+u,t+w)$ for large $\|\beta\|$.
Hence, we have the conditions to apply the result of \citet
{Fors2010} in each orthant
to guarantee a finite integral.


Now we\vspace*{1pt} require to compute the marginal likelihood induced by
$p^{\mathrm{IM}}(\beta\g h, t)$.
This is given by
%
\begin{eqnarray}
\label{marginalIM}\quad
&&
m^{\mathrm{IM}}(y \g h, t) \nonumber\\
&&\quad= \Biggl\{ \prod
_{i=1}^N \pmatrix{n_i
\cr
y_i} \Biggr\} \int_{\Re^{k+1}} L(\beta\g y,n)
p^{\mathrm{IM}}(\beta\g h,t)\,d\beta
\nonumber\\[-8pt]\\[-8pt]
&&\quad= 
\Biggl\{ \prod
_{i=1}^N \pmatrix{n_i
\cr
y_i} \Biggr\} \nonumber\\
&&\qquad{}\cdot\sum_x
m_0(x) \frac{Q(x+u+y,t+w+n,h)} {Q(x+u,t+w,h)}.\nonumber
\end{eqnarray}
In practice, we need an efficient method to compute $Q(z+x,s+t,h)$ for
$(z,s)=(u,w)$
and $(z,s)=(u+y,w+n)$.
Since
\begin{eqnarray*}
&&
Q(z+x,s+t,h)\\
&&\quad= \int_{\Re^{k+1}} \frac{L(\beta\g z,s)L(\beta\g x,t)}{Q(z,s,0)} Q(z,s,0) \\
&&\hspace*{24.2pt}\qquad{}\cdot\Biggl
\{\prod_{j=1}^k \beta_j^{2h}
\Biggr\} \,d\beta
\\
&&\quad= Q(z,s,0) \vat^{p^C(\beta\g z,s)} \Biggl\{ \Biggl(\prod
_{j=1}^k \beta_j^{2h} \Biggr)
L(\beta\g x,t) \Biggr\},
\end{eqnarray*}
one can simulate from the conjugate local prior $p^C(\beta\g\break z,s)$
using MCMC methods
and obtain $m^{\mathrm{IM}}(y \g h, t)$ as a mixture, with respect to $m_0(x)$,
of ratios of expectations; the normalizing constants $Q(z,s,0)$ for $h=0$
will be computed once and for all, for a given data set, again using
MCMC methods; see Section~\ref{subseclogitModels}.


\section{Applications}
\label{secappl}


In this section we apply our methodology to two problems.
The first application concerns a set of randomized trials
and uses results presented in Section~\ref{sectwoBinomial};
the second application performs model selection within a logistic
regression framework
to analyze the relationship between the probability of patients' survival
and two binary covariates,
and makes use of results presented in Section~\ref{seclogreg}.

\subsection{Randomized Trials}
\label{subsecrandomized}

We analyze data from 41 randomized trials
of a new surgical treatment for stomach ulcers.
For each trial
the number of occurrences and nonoccurrences
under Treatment (the new surgery, \emph{group 1}) and
Control (an older surgery, \emph{group 2}) are reported;
see Efron (\citeyear{Efro1996}, Table 1).
Occurrence here refers to an adverse event: recurrent bleeding.
\citet{Efro1996} analyzed these data
with the aim of performing a meta-analysis, using empirical Bayes methods.
On the other hand, our objective is to establish
whether the probability of occurrence is the same
under Treatment and Control in each individual table;
for a similar analysis see \citet{CaseMore2009}.
We base our analysis on the intrinsic moment priors
of Section~\ref{sectwoBinomial}, letting $b_0=1/2$ and comparing the results
given by different choices of $h$ and~$t$. Specifically,
we perform a sensitivity analysis with respect to the actual choice of $t$,
and a cross-validation study of the predictive performance
achieved by different choices of $h$.

\subsubsection{Sensitivity analysis}

Recall that a crucial hyperparameter is represented
by the overall training sample size $t_+$,
which is then further split into the two groups, $t_+=t_1+t_2$.
On the basis of our study of the characteristics of intrinsic moment priors
for the comparison of two proportions,
we suggest a sensitivity
analysis with $t_+>t_+^*(h)$.
Specifically, we here let $t_+$ vary from $t_+^*(h)$ to $t_+^{**}(h)=t_+^*(h+1)$,
where $h=0$ (standard local prior) or $h=1$ (recommended nonlocal prior);
recall that $t_+^*(0)=0$, $t_+^*(1)=8$ and $t_+^*(2)=14$.
We choose $t_1$ and $t_2$ approximately proportional to
the trial sample sizes for Treatment and Control, $n_1$ and $n_2$,
and $b_1$ and $b_2$ exactly proportional to these quantities,
with $b_1+b_2=b_0=1/2$ (unit prior information).
For all the above pairs $(h, t)$, and all 41 tables in the data set,
we evaluate the posterior probability of the null model.
%

\begin{figure*}

\includegraphics{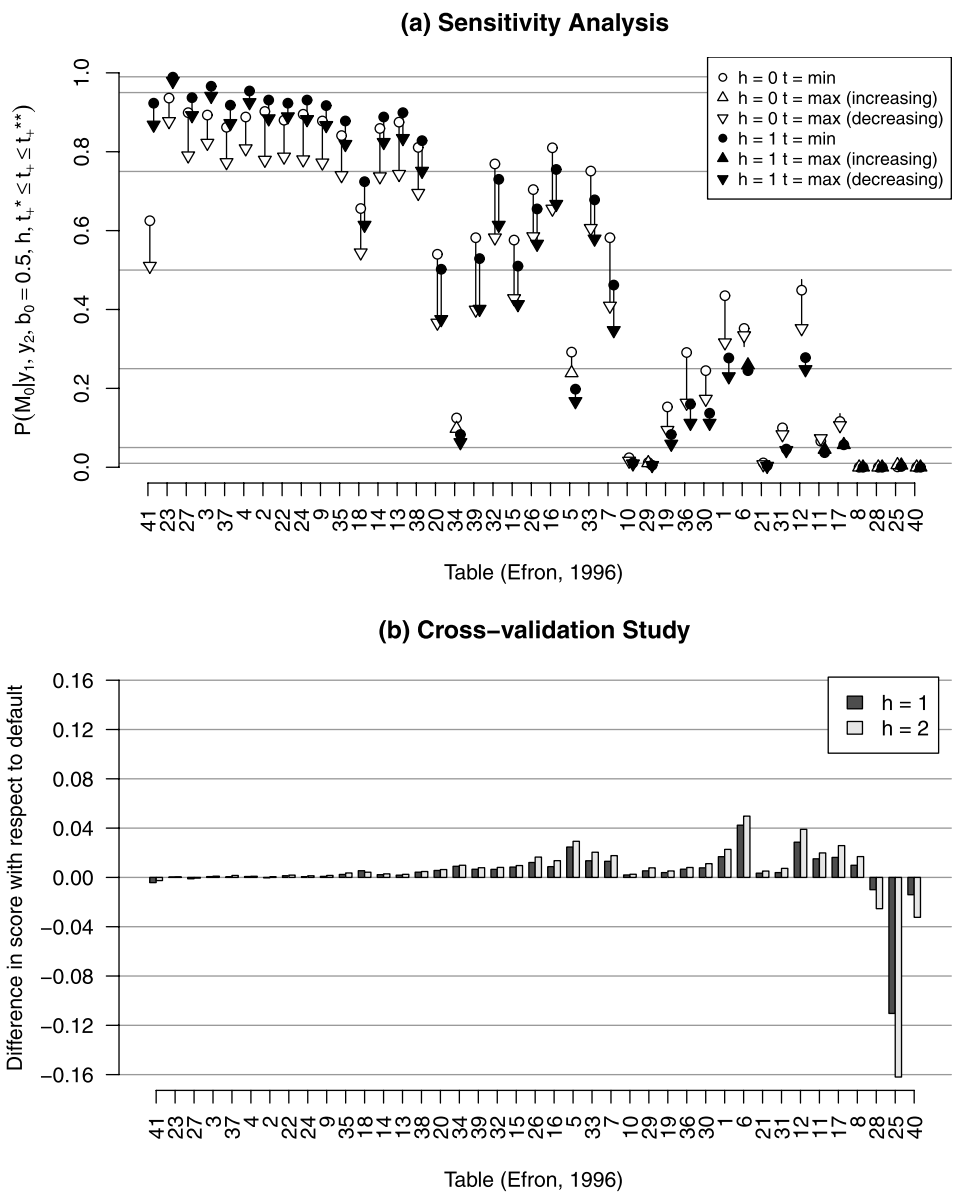}

\caption{Results of sensitivity analysis and cross-validation study:
each number on the horizontal axis identifies a table.}
\label{figApplication}\vspace*{6pt}
\end{figure*}

We report our findings
in Figure~\ref{figApplication}(a), where the tables are arranged
(for a better appreciation of our results) from left\vspace*{1pt} to right
in increasing order of $|\frac{y_1}{n_1}-\frac{y_2}{n_2}|$
(absolute difference in observed fractions):
this explains the mostly declining pattern of the posterior probabilities
of the null model.
The range of these probabilities is depicted as a vertical segment,
separately for the standard intrinsic and the intrinsic moment prior,
and the values for $t=t^*$ and $t=t^{**}$ are marked with
circles and triangles, respectively, so that in most cases
(thanks to a monotonic behavior) we can see an arrow describing
the overall change in probability.
One can identify three sets of tables: left-hand (up to Table 38),
center (tables from 20 to 7) and right-hand (remaining tables).
Some specific comments follow below.

Consider first the left-hand tables.
Except for Table 41 under the local prior
(and possibly Table 18), the posterior probability of ${\cal M}_0$
ranges well above the value $0.5$,
which can be regarded as a conventional decision threshold for model choice
under a $\{0,1\}$-loss function.
The nonlocal intrinsic moment prior (black triangle) produces
values for the posterior probability of ${\cal M}_0$ higher than
under the standard intrinsic prior (white triangle):
this is only to be expected, because of the nonlocal versus local
nature of these priors.
The effect is dramatic for Table 41,
which is characterized by counting no occurrences at all.
All arrows point downward:
this is the effect of the intrinsic procedure;
when the data support the null model, the action of pulling the prior toward
the null subspace makes the alternative more competitive and takes evidence
away from~${\cal M}_0$.
For this first group of tables,
a robust conclusion can be reached
in favor of the equality of proportions between the two groups.
Next consider the tables in the center.
Four of these tables (20, 39, 15, 7) exhibit a posterior probability of
the null
hovering over the $0.5$ threshold, so that no robust conclusion can be drawn;
four of them (32, 26, 16, 33) give a robust conclusion in favor of the null,
while two of them (34, 5) give a robust conclusion against the null.
Leaving aside these last two tables, which are characterized by zero occurrences
in one of the two groups and are more similar in behavior
to the right-hand tables, all arrows point downward,
indicating that here too the intrinsic procedure is working in favor of
the alternative.
Notice that now the local priors give more credit to the null than the
nonlocal priors,
showing that the steeper evidence gradient of the latter gets into play.
Finally, the pattern of the right-hand tables indicates a low support
for the null,
with the possible exception of Tables 1 and 12.
Ranges become shorter, and on some occasions negligible,
especially for the nonlocal priors. Some arrows point upward:
this is the action of the intrinsic procedure in favor of ${\cal M}_0$,
because the data do not support the null model;
the two Bayesian models are on the way of becoming equivalent.
For the tables in this last group,
a robust conclusion against the null can be drawn.

\subsubsection{Cross-validation study}

We now compare the predictive performance
of the intrinsic moment priors with $h=0$, $h=1$ and $h=2$,
taking for granted that $t_+$ should be equal to $t_+^*$
(for any given value of $h$)
and $t_1$ and $t_2$ should be (approximately) proportional to $n_1$ and $n_2$.
To this aim, we assign a logarithmic score to each probability
forecast
$p$, say,
of an event $E$: the score is $\log(p)$, if $E$ occurs, and $\log(1-p)$,
if $\bar{E}$ occurs; this is a proper scoring rule (\cite{BernSmit1994},
Section 2.7.2).
Notice that each score is negative, the maximum value it can achieve is zero,
and higher scores indicate a better prediction.
Suppose we want to predict the outcome for a patient who is an
occurrence in group~1.
We exclude this patient from the data set and compute the probability for
an occurrence of such a patient,
$\hat{\theta}_1^{(1)}$, as the Bayesian model average of the
posterior means of $\theta_1$ under $\mathcal{M}_1$
and $\theta$ under $\mathcal{M}_0$
based on counts
$(y_1-1,n_1-y_1,y_2,n_2-y_2)$;
similarly, to predict the outcome for a patient who is an occurrence in
group 2,\vspace*{1pt}
we compute her probability of occurrence, $\hat{\theta}_2^{(1)}$,
upon interchanging subscript 1 and 2 above.
On the other hand, to predict the outcome for a patient who is a
nonoccurrence in group~1,
we compute the corresponding probability of an occurrence, $\hat
{\theta}_1^{(0)}$,
as the Bayesian model average of the
posterior means of $\theta_1$ under $\mathcal{M}_1$ and
$\theta$ under $\mathcal{M}_0$ based on counts
$(y_1,n_1-y_1-1,y_2,n_2-y_2)$; as before,
the computation of $\hat{\theta}_2^{(0)}$,
for a patient who is a nonoccurrence in group 2,
requires interchanging subscripts 1 and 2.
In the spirit of cross-validation, we repeat the analysis for each\vadjust{\goodbreak} patient
and compute the overall mean score
\begin{eqnarray*}
S &=& \bigl(y_1\log\hat{\theta}_1^{(1)}+ (n_1-y_1)\log\bigl(1-\hat{\theta}_1^{(0)}\bigr)\\
&&\hspace*{2pt}{}+
y_2\log\hat{\theta}_2^{(1)}+ (n_2-y_2)\log\bigl(1-\hat{\theta}_2^{(0)}\bigr)
\bigr)\\
&&{}/({n_1+n_2}).
\end{eqnarray*}
%

Now let $S_h$ be the score associated with the intrinsic moment prior
of order $h$,
$h=0,1,2$. Of particular interest are the differences $S_1-S_0$ and $S_2-S_0$.
A positive value for $S_1-S_0$, say, means that the prior with $h=1$ produces
on average a better forecasting system than
the standard intrinsic prior ($h=0$);
notice that the latter coincides with the default prior because $t_+^*=0$.
One can use a first order expansion of the logarithmic score
to gauge the difference more concretely:
a positive difference $S_1-S_0=d>0 $ means
that the prior with $h=1$ generates
``correctly-oriented probability forecasts''
(higher values for occurrences and lower values for nonoccurrences)
which are, on average, $d \times100 \%$ better than
those produced by the standard intrinsic prior.
Here the average is taken over the combination of event outcomes
(occurrence/nonoccurrence) and groups (Treatment/Control)
with weights given by the observed sample frequencies.
Since $d>0$ is an average of score differences over the four blocks of events,
there is no guarantee of a uniform improvement in prediction across all
of them.

Figure~\ref{figApplication}(b) reports the results of
our cross-vali\-dation study with the tables again arranged
from left to right in increasing order of
absolute difference in observed fractions.
Essentially for all tables,
but with the notable exception of the last three,
the nonlocal intrinsic moment priors perform better than
the standard intrinsic prior, with differences in score
ranging from $-0.42\%$ to $4.2\%$ (median improvement $0.54\%$) when $h=1$
and from $-0.26\%$ to $5.0\%$ (median improvement $0.68\%$) when $h=2$.
On the other hand, for the last three tables, which are clearly against
the null,
the performance of nonlocal priors is much worse:
this happens because the intrinsic moment priors produce a greater
degree of posterior shrinkage toward the null within the alternative model.
Differences in score range from $-1.0\%$ down to $-11\%$, when $h=1$,
and from $-2.5\%$ down to $-16\%$, when $h=2$.
Notice that
the intrinsic moment prior predicts better with $h=2$ than with $h=1$
when the difference in score is positive, but the reverse occurs for negative
differences in score; in the latter case the performance can be
appreciably worse.
On grounds of prudence, these results seem to reinforce
our recommendation in favor of the choice $h=1$.

\subsection{Logistic Regression Models for Survival Data}
\label{subseclogitModels}

In Table~\ref{tablogitData} we consider a data set previously examined
in \citet{DellaEtal2002};
see also references therein for further analyses of the same problem.
Our aim is to investigate the relationship between the probability of
Survival, on the one hand, and two binary covariates: Severity of condition
and Antitoxin medication.

\begin{table}
\caption{Survival data}
\label{tablogitData}
\begin{tabular*}{\tablewidth}{@{\extracolsep{\fill}}lccc@{}}
\hline
\textbf{Condition} & \textbf{Antitoxin} & \textbf{Death} & \textbf{Survival} \\
\hline
More severe & Yes & 15 & \hphantom{0}6 \\
& No & 22 & \hphantom{0}4 \\
Less severe & Yes & \hphantom{0}5 & 15 \\
& No & \hphantom{0}7 & \hphantom{0}5 \\
\hline
\end{tabular*}
\end{table}

The full model is given by
\begin{eqnarray*}
y_{jl} \g\theta_{jl} &\stackrel{\mathrm{ind}} \sim& 
\operatorname{Bin}(n_{jl}, \theta_{jl} ),
\\
\log \biggl( \frac{\theta_{jl}}{1 - \theta_{jl}} \biggr) &=& \alpha+ \beta_j +
\gamma_l + \delta_{jl},
\end{eqnarray*}
$j,l = 1, 2$, where $y_{jl}$, $n_{jl}$ and $\theta_{jl}$ are the
number of survivals,
the total number of patients and the probability of survival under
level $j$ of Severity and level $l$ of Antitoxin medication;
$\alpha$, $\beta_j$, $\gamma_l$ and $\delta_{jl}$ are the model parameters
corresponding to the intercept, Severity effect,
Antitoxin effect, and interaction effect of Severity and Antitoxin.
The number of free parameters is actually four:
intercept, two main effects and one interaction.

\begin{table*}
\tablewidth=360pt
\caption{Posterior probabilities of five logistic regression models
for the survival data in Table \protect\ref{tablogitData},
using intrinsic moment priors
with total weight of evidence on corresponding minimal data given in the last column.
Each model is described through the main effect(s) it includes beside
intercept}
\label{tablogit}
\begin{tabular*}{\tablewidth}{@{\extracolsep{\fill}}ld{2.0}cccccd{2.0}@{}}
\hline
$\bolds{h}$ & \multicolumn{1}{c}{$\bolds{t_+}$} &
\textbf{Intercept-only} & \textbf{Severity} & \textbf{Antitoxin}
& \textbf{Sever}${}\bolds{+}{}$\textbf{Antitox} & \textbf{Full model} &
\multicolumn{1}{c@{}}{\textbf{TWOE}}\\
\hline
0 & 0 & 0.01 & 0.61 & 0.01 & 0.35 & 0.02 & 7 \\
& 4 & 0.01 & 0.56 & 0.01 & 0.40 & 0.01 & 4 \\
& 8 & 0.00 & 0.44 & 0.01 & 0.51 & 0.03 & 3 \\
& 12 & 0.00 & 0.35 & 0.01 & 0.54 & 0.10 & 2 \\
& 16 & 0.00 & 0.33 & 0.01 & 0.54 & 0.12 & 2 \\
& 20 & 0.00 & 0.29 & 0.01 & 0.53 & 0.17 & 2 \\
& 24 & 0.00 & 0.26 & 0.01 & 0.52 & 0.21 & 2 \\
1 & 0 & 0.22 & 0.77 & 0.01 & 0.00 & 0.00 & 2 \\
& 4 & 0.03 & 0.86 & 0.01 & 0.10 & 0.00 & 7 \\
& 8 & 0.01 & 0.85 & 0.01 & 0.13 & 0.00 & 6 \\
& 12 & 0.00 & 0.67 & 0.01 & 0.31 & 0.00 & 6 \\
& 16 & 0.00 & 0.62 & 0.01 & 0.36 & 0.00 & 6 \\
& 20 & 0.00 & 0.52 & 0.01 & 0.46 & 0.00 & 6 \\
& 24 & 0.00 & 0.45 & 0.01 & 0.53 & 0.00 & 5 \\
2 & 0 & 0.95 & 0.05 & 0.00 & 0.00 & 0.00 & -2 \\
& 4 & 0.13 & 0.86 & 0.01 & 0.01 & 0.00 & 21 \\
& 8 & 0.03 & 0.95 & 0.01 & 0.00 & 0.00 & 25 \\
& 12 & 0.01 & 0.93 & 0.01 & 0.05 & 0.00 & 26 \\
& 16 & 0.01 & 0.89 & 0.01 & 0.09 & 0.00 & 27 \\
& 20 & 0.00 & 0.80 & 0.01 & 0.19 & 0.00 & 26 \\
& 24 & 0.00 & 0.70 & 0.01 & 0.29 & 0.00 & 25 \\
\hline
\end{tabular*}
\end{table*}

We are interested in five distinct logistic regression models:
the intercept-only model, two models with a single main effect each
(plus intercept),
one model with two additive main effects (plus intercept),
and the full model. We wish to compare them through their posterior
probabilities
based on our intrinsic moment priors.
Our results are summarized in Table~\ref{tablogit},
where we report posterior model probabilities with an accuracy
(standard error)
of approximately 0.01.

Computations were performed using the methodology presented in
Section~\ref{seclogreg}.
In particular, for the choice of prior hyperparameters, we used formula
(\ref{eqwiAndui})
with $w_{+}=1$. A uniform prior on the model space was assumed.
For each model, a random walk Metropolis--Hastings sampler was
implemented\break
through the function \texttt{metrop()} of
the \texttt{R} package \texttt{mcmc} (\cite{Geye2010}).
Prior and posterior normalizing constants with $h=0$ were computed,
once and for all,
using the method by \citet{ChibJeli2001} on chains of length 40,000
after thinning by a factor 20;
the proposal distributions were tuned so as to obtain
acceptance rates between 24\% and 28\%.
Different chains with the same features were used
to compute the ratios of posterior expectations needed to calculate
the intrinsic moment marginal likelihood (\ref{marginalIM})
as a mixture with respect to $m_0(x)$.
Since the mixing step proved to be computationally demanding,
we used C (within R).

Differently from the case of the comparison of two proportions,
for general logistic regression models there seems to be no
simple method to determine $t^*_+$ once and for all,
because the explanatory variables are different in each application.
Moreover, extending the methodology of total weight of evidence
presented in Section~\ref{subsecChoosingTrainingSampleSize} to the case
of more than two models appears to be nontrivial.
In the present application we found it natural to let $n_{jl}\equiv1$
for minimal data, and contented ourselves with computing
the total weight of evidence for the full model against the
intercept-only one,
focusing on the two models farthest from each other.
We used this information to guide our choice of $t_+$
in the context of a sensitivity analysis across a grid of values
for the hyper\-parameters $h=0,1,2$ and $t_+=0,4,8,12,16,20,24$;
the actual values of $t_{jl}$ were obtained, by rounding them,
as approximately proportional to $n_{jl}$.
In general, the choice of $t$ could depend on the model,
which could help comparing models of very different dimension,
but in the present case we avoided this additional complexity.

The values of the total weight of evidence in Table~\ref{tablogit}
suggest that we should take $t_+=0$ when $h=0$, $t_+=4$ when $h=1$,
and $t_+=16$ when $h=2$. However, the last column of Table~\ref{tablogit}
is not stable across different MCMC runs,
and it should be considered as merely indicative.
This is not surprising, because the total weight of evidence
was quite flat around its maximum in both Figure~\ref{figbernoullisecond}(b)
and Figure~\ref{figCharact}(a); it is a problem that cannot be solved
by a feasible increase in chain length. The clear message appears to be that
$t_+=12$ is too much when $h=0$, $t_+=0$ is not a good choice when $h=1$,
and $t_+=4$ is not enough when $h=2$.
Notice that the first value would give some credit to the full model,
while the last two values would attribute a sizeable posterior probability
to the intercept-only model.
Then, if a recommended value $t^*_+(h)$
has to be singled out for each value of~$h$,
the choice $t^*_+(0)=0$, $t^*_+(1)=8$ and $t^*_+(2)=16$
achieves a better scaling with respect to~$h$,
and it is in line with the values found for the comparison of two proportions.
Here too the intrinsic step appears to be necessary for nonlocal priors only.

Bearing in mind that the intercept term is present in each model,
the posterior model probabilities reported in Table~\ref{tablogit}
suggest that
the two models ``Severity'' and ``Severity${}+{}$Antitoxin'' account for
at least 90\% of the probability mass in all reasonable scenarios.
Specifically, model ``Severity'' is a clear winner under the nonlocal priors,
except for $h=1$ and the highest values of $t_+$, which are far from
$t^*_+(1)$.
The situation is more mixed under the local priors:
the leadership of ``Severity'' is not equally clear,
and it fades away as $t_+$ increases;
these results are in line with those obtained by \citet{DellaEtal2002}
using several MCMC schemes all based on local normal parameter priors.
While local and nonlocal priors broadly agree
on the two leading models,
they diverge on the allocation of probability mass between them:
for values of $t_+$ close to the recommended ones
nonlocal priors more sharply select the parsimonious model
``Severity'',
dropping its more complex competitor ``Severity${}+{}$Antitoxin''.


\section{Discussion}
\label{secdisc}

In this paper we have presented a general approach to
objective Bayesian testing for nested hypotheses in discrete data models.
The only required input is a default (proper) parameter prior under
each of the entertained models. Next, a default nonlocal prior is
derived, and finally a procedure based on the intrinsic
methodology is applied.
The fundamental tool in our approach is represented by a particular
class of
nonlocal priors, which we name intrinsic moment priors. These
distributions combine
the virtues of nonlocal priors and intrinsic priors to obtain balanced
objective tests,
whose learning rate is improved (strongly accelerated
when the smaller model holds) relative to current local prior methods, while
their small sample evidence
is broadly comparable with that afforded by modern objective methods,
including those based on intrinsic priors.

An important feature of intrinsic moment priors is represented by
the training sample size. We handle the choice of this hyperparameter
in a novel way,
and quite differently from current intrinsic approaches,
using the notion of total weight of evidence.
This criterion
looks promising, at least for finitely discrete data models,
but it cannot be naively extended to the countably infinite or
continuous case
because we cannot weight data values uniformly;
a suitable weighting data measure should be devised,
whose choice, however, remains an open issue.
Whether or not an optimal value for the training sample size can be
found,
one can always carry out a sensitivity analysis,
an exercise we typically recommend to assess robustness of conclusions
with respect to this hyperparameter.

Our approach for the construction of prior distributions is based on
a comparison of two nested models. When several models are entertained,
we select the null model as a natural baseline,
because it is nested within any other model,
similarly to methods based on intrinsic priors (\cite{GiroEtAl2006})
or on mixtures of $g$-priors (\cite{LianPauletal2008}).
This choice of course can be modified, if an alternative minimal model
is available.
Clearly, the baseline model acquires a special status in this approach.
Assignments of parameter priors for pairwise comparison of models
which are symmetric in nature, because they do not require a baseline model,
are developed in \citet{CanoEtAl2008}.

While our analysis was solely based on proper priors, we emphasize
that moment priors can also be improper. The subsequent analysis can
then proceed through an intrinsic step, as in this paper, or through
other methods currently available to deal with improper priors for
model comparison, such as expected posterior priors, or fractional
Bayes factors (\cite{Ohag1995}); for an application of the latter
methodology see
\citet{ConsLaro2011} and \citet{AltoConsLaRo2011}.

\begin{appendix}\label{app}
\section*{Appendix: Bayes Factor Asymptotics}

We study the asymptotic learning rate of BFs for comparing nested models,
allowing for nonlocal parameter priors, under fairly general assumptions.
It will be understood that the data be discrete,
but this assumption will not be crucial to our results.
We start with the notion of a regular, possibly misspecified, model.
%
\begin{defin}
\label{defregmod}
A nonsingleton statistical\break model $\mathcal{M}=\{f^n(\cdot|\xi),\xi
\in\Xi\}$ for
a sequence of\vadjust{\goodbreak} data $y^{(n)}=(y_1,\ldots,y_n)$ taking values in
$\mathcal{Y}^n$
is regular with respect to the strictly positive sampling density
$q^n(\cdot)$
if the following assumptions hold:
\begin{enumerate}
\item$\Xi$ is an open subset of $\Re^d$;
\item$f(\cdot|\xi)$ is strictly positive, for all $\xi\in\Xi$;
\item the Kullback--Leibler projection of $q(\cdot)$ on $\mathcal{M}$
is well-defined,
that is, there exists a unique $\xi^\star\in\Xi$ such that
$K_q(\xi^\star)=\inf_{\xi\in\Xi}K_q(\xi)$,
where $K_q(\xi)=\vat_q\log\{q(y_1)/f(y_1|\xi)\}$
is the Kullback--Leibler divergence from $q(\cdot)$ to $f(\cdot|\xi)$;
\item$\var_q\log{\{q(y_1)/f(y_1|\xi^\star)\}}<\infty$;
\item the unit log-likelihood $\ell(y_1|\cdot)=\log f(y_1|\cdot)$ is
twice continuously differentiable on $\Xi$
with gradient\break $s(y_1|\cdot)$ and Hessian matrix $H(y_1|\cdot)$,
for all $y_1\in\mathcal{Y}$;
\item$\vat_q|\ell(y_1|\xi^\star)|<\infty$ and $\vat_q\|s(y_1|\xi
^\star)\|_2^2<\infty$;
\item there exist a spheric neighborhood $B$ of $\xi^\star$,
$B\subseteq\Xi$,
and a function $c(\cdot)$ from $\mathcal{Y}$ to $\Re_+$, with\break $\vat
_q c(y_1)<\infty$,
such that $\sup_{\xi\in B}\|H(y_1|\xi)\|_\infty\le\break c(y_1)$;
\item the upper level sets of the average log-likelihood
$\bar{\ell}_n(y^{(n)}|\cdot)=n^{-1}\sum_{i=1}^n\ell(y_i|\cdot)$,
that is, all sets of the form $\{\xi\in\Xi\dvtx \bar{\ell
}_n(y^{(n)}|\xi)>\lambda\}$,
with $\lambda\in\Re$, are connected, for all $y^{(n)}\in\mathcal{Y}^n$.
\end{enumerate}
A singleton model $\mathcal{M}=\{f^n(\cdot|\xi^\star)\}$ is regular
with respect to $q^n(\cdot)$ if 2 and 4 hold, being understood that
$K^\star=K_q(\xi^\star)<\infty$ and $\Xi=\{\xi^\star\}$.
\end{defin}
For a nonsingleton model, the Kullback--Leibler divergence $K^\star$
from $q(\cdot)$ to $\mathcal{M}$ is necessarily finite;
otherwise $\xi^\star$ would not be uniquely defined.
If $q(\cdot)\in\mathcal{M}$, then assumption 3 is implied by identifiability,
while assumption 4 is trivial, because $f(\cdot|\xi^\star)=q(\cdot)$.
On the other hand, if $q(\cdot)\notin\mathcal{M}$, then $K^\star>0$,
because $K_q(\xi)=0$ implies $q(\cdot)=f(\cdot|\xi)$.

Assumption 7 in Definition~\ref{defregmod} implies
$\vat_q\|H(y_1|\break\xi)\|_\infty<\infty$ for all $\xi\in B$
and can be extended to the unit score vector
by writing the Taylor expansion with integral remainder
\begin{eqnarray*}
s(y_1|\xi)&=&s\bigl(y_1|\xi^\star\bigr)\\
&&{}+\int
_{0}^1 H\bigl(y_1|
\xi^\star+t\bigl(\xi-\xi ^\star\bigr)\bigr) \bigl(\xi-
\xi^\star\bigr)\,dt, %
\end{eqnarray*}
which gives
$\sup_{\xi\in B}\|s(y_1|\xi)\|_\infty\le
\|s(y_1|\xi^\star)\|_2+\break d c(y_1)\sup_{\xi\in B}\|\xi-\xi^\star\|
_2=b(y_1)$;
this in turn implies $\vat_q\|s(y_1|\xi)\|_2<\infty$ for all $\xi
\in B$.
Similarly, assumption 7 in Definition~\ref{defregmod}
can be extended to the unit log-likelihood, obtaining
\begin{eqnarray*}
\sup_{\xi\in B}\bigl|\ell(y_1|\xi)\bigr|&\le& \bigl|\ell
\bigl(y_1|\xi^\star\bigr)\bigr|+d b(y_1)\sup
_{\xi\in B}\bigl\|\xi-\xi^\star\bigr\|_2\\
&=&a(y_1)
\end{eqnarray*}
and $\vat_q|\ell(y_1|\xi)|<\infty$ for all $\xi\in B$.
We are now ready for a technical lemma.
%
\begin{lemma}
If a nonsingleton statistical model $\mathcal{M}=\{f^n(\cdot|\xi
),\xi\in\Xi\}$
is regular with respect to the strictly positive sampling density
$q^n(\cdot)$,
the expected log-likelihood $L_q(\xi)=\vat_q\ell(y_1|\xi)$, $\xi
\in B$,
is twice continuously differentiable on $B$ with gradient
$L_q^\prime(\cdot)=\vat s(y_1|\cdot)$ and Hessian matrix
$L^{\prime\prime}_q(\cdot)=\vat_q H(y_1|\cdot)$.
\end{lemma}
\begin{pf}
$L_q(\cdot)$ is differentiable on $B$ with gradient $L^\prime_q(\cdot
)=\vat_q s(y_1|\cdot)$
because assumption 7 extended to the unit score vector allows the
derivative to pass
under the integral sign; see, for instance, the lemma on page 124 of
\citet{Ferg1996}.
In the same way, it follows from assumption 7 that the expected score vector
$L^\prime_q(\cdot)$ is differentiable on $B$
with derivative matrix $L^{\prime\prime}_q(\cdot)=\vat_q
H(y_1|\cdot)$.
Then,\vspace*{1pt} through a direct application of Lebsegue's Dominated Convergence theorem,
it also follows from assumption 7 that $L^{\prime\prime}_q(\cdot)$
is continuous on~$B$.
\end{pf}
Since $K_q(\xi)=K_q(\xi^\star)+L_q(\xi^\star)-L_q(\xi)$, for $\xi
\in B$,
with $K_q(\xi^\star)<\infty$,
the Kullback--Leibler divergence from $q(\cdot)$ to $f(\cdot|\xi)$
is also twice continuously differentiable on $B$, as a function of $\xi$,
with gradient $K^\prime_q(\cdot)=-L^\prime_q(\cdot)$
and Hessian matrix $K^{\prime\prime}_q(\cdot)=-L^{\prime\prime
}_q(\cdot)$.
Then, since $\xi^\star$ is the unique minimum of $K_q(\cdot)$ on~$\Xi$,
$\xi^\star$ is also the unique maximum of $L_q(\cdot)$ on $B$.
Finally, since $\xi^\star$ is an interior point of $B$, we find
$L_q^\prime(\xi^\star)=K_q^\prime(\xi^\star)=0$
and $L_q^{\prime\prime}(\xi^\star)$ a negative definite matrix,
equivalently, $K_q^{\prime\prime}(\xi^\star)$ a positive definite matrix.

We now give a classical theorem on maximum likelihood asymptotics,
but for a possibly misspecified model; see Theorems 17 and 18 of
\citet{Ferg1996}.

\begin{theorem}\label{thmmle}
Let $y^{(n)}=(y_1,\ldots,y_n)$ be data arising under i.i.d. sampling
from a distribution with strictly positive density $q^n(\cdot)$
and $\mathcal{M}=\{f^n(\cdot|\xi),\break\xi\in\Xi\}$ be a nonsingleton
model for such data,
which we assume to be regular with respect to $q^n(\cdot)$. Then,
there exists $\hat\xi_n$ such that, almost surely, for large enough $n$,
$\hat\xi_n$ is a global maximum of the log-likelihood.
Moreover, for any such maximum likelihood estimator $\hat\xi_n$,
the following conditions are satisfied:
\begin{longlist}[(iii)]
\item[(i)] almost surely, for large enough $n$,
$\hat\xi_n$ is a root of the score equation, that is,
$\bar{s}_n(y^{(n)}|\hat\xi_n)=0$,
where $\bar{s}_n(y^{(n)}|\cdot)=n^{-1}\sum_{i=1}^n s(y_i|\cdot)$ is
the average score;
\item[(ii)] almost surely $\hat\xi_n\to\xi^\star$, as $n\to
\infty$;
\item[(iii)] for all small enough $\rho>0$ there exists $\delta>0$
such that,
almost surely, for large enough $n$, it holds that
\[
\sup_{\xi\in\Xi\cap\{\|\xi-\xi^\star\|_2\ge\rho\}} \bar\ell_n\bigl(y^{(n)}|\xi
\bigr)<\bar\ell_n\bigl(y^{(n)}|\hat\xi_n\bigr)-
\delta; %
\]
\item[(iv)] $n^{1/2}(\hat\xi_n-\xi^\star)\rightsquigarrow\mathcal
{N}_d(0,V^\star)$,
as $n\to\infty$, where
\begin{eqnarray*}
V^\star&=& L^{\prime\prime}_q\bigl(\xi^\star\bigr)^{-1} \vat_q\bigl\{s
\bigl(y_1|\xi^\star\bigr)s\bigl(y_1|
\xi^\star\bigr)^\top\bigr\}\\
&&{}\cdot L^{\prime\prime}_q
\bigl(\xi^\star\bigr)^{-1}. %
\end{eqnarray*}
\end{longlist}
Finally, as a consequence of \textup{(iv)}, we can write
$\hat\xi_n-\xi^\star=\mathcal{O}_p(n^{-1/2})$.
\end{theorem}
\begin{pf}
Let $S$ be a compact sphere contained in $B$
and denote by $\hat\xi_n$ a maximum likelihood estimator
of $\xi$ constrained to $S$, which always exists
because the average log-likelihood $\bar\ell_n(y^{(n)}|\cdot)$
is continuous on $S$. Then, fix $\rho>0$ small enough for
$C=\{\xi\in S\dvtx \|\xi-\xi^\star\|_2\ge\rho\}$ to be nonempty.
Since $C$ is compact, a uniform version of the Strong Law of Large Numbers
for continuous dominated summands (\cite{Ferg1996}, Theorem~16) gives
$\sup_{\xi\in C}\llvert \bar\ell_n(y^{(n)}|\xi)-L_q(\xi)\rrvert \to0$,
as $n\to\infty$, almost surely.
Now $\sup_{\xi\in C}L_q(\xi)=\break L_q(\xi^\star)-3\delta$,
for some $\delta>0$, because $\xi^\star\notin C$ and $L_q(\cdot)$
is continuous on $C$.
Hence, almost surely, for large enough $n$, we have
$\sup_{\xi\in C}\bar\ell_n(y^{(n)}|\xi)<L_q(\xi^\star)-2\delta$.
However, due to the ordinary Strong Law of Large Numbers,
we also have $\bar\ell_n(y^{(n)}|\xi^\star)>L_q(\xi^\star)-\delta$.
Then, the connected set
$\{\xi\in\Xi\dvtx \bar\ell_n(y^{(n)}|\xi)>L_q(\xi^\star)-2\delta\}$
contains $\xi^\star$ but has empty intersection with $C$.
Since $C\neq\varnothing$,
this upper level set has also empty intersection with $\Xi\setminus S$.
It follows that $\hat\xi_n$ is a global maximum of the log-likelihood.

Now let $\hat\xi_n$ be any global maximum likelihood estimator.
Since $\Xi$ is open, $\hat\xi_n$ is necessarily an interior point of
$\Xi$
and (i) follows. Moreover, the above argument shows that
$\|\hat\xi_n-\xi^\star\|_2<\rho$, for any small enough $\rho>0$,
which is enough to prove (ii). The above argument also gives
\begin{eqnarray*}
\sup_{\xi\in\Xi\cap\{\|\xi-\xi^\star\|_2\ge\rho\}}\bar\ell _n\bigl(y^{(n)}|\xi
\bigr) &\le& L_q\bigl(\xi^\star\bigr)-2\delta\\
&<&\bar
\ell_n\bigl(y^{(n)}|\hat\xi_n\bigr)-\delta,
\end{eqnarray*}
which is (iii). Therefore, only (iv) remains to be shown.

Consider the Taylor expansion with integral remainder
\[
\bar{s}_n\bigl(y^{(n)}|\xi^\star\bigr)= \int
_0^1\bar{H}_n
\bigl(y^{(n)}|\hat\xi_n+t\bigl(\xi^\star-\hat
\xi_n\bigr)\bigr) \bigl(\xi ^\star-\hat\xi_n
\bigr)\,dt, %
\]
where $\bar{H}_n(y^{(n)}|\cdot)=n^{-1}\sum_{i=1}^nH(y_i|\cdot)$
is the average Hessian of the log-likelihood and
we have used\vadjust{\goodbreak} the fact that $\hat\xi_n$ is a root of the score equation.
The central limit theorem tells us that
$\bar{s}_n(y^{(n)}|\xi^\star)\rightsquigarrow
\mathcal{N}_d(0,\break\vat_q\{s(y_1|\xi^\star)s(y_1|\xi^\star)^\top\})$,
as $n\to\infty$. Hence, Slutsky's theorem will give us (iv) if we
show that
\begin{eqnarray*}
R_n\bigl(\hat\xi_n,\xi^\star\bigr)&=& \int
_0^1\bar{H}_n
\bigl(y^{(n)}|\hat\xi_n+t\bigl(\xi^\star-\hat
\xi_n\bigr)\bigr)\,dt \\
&\to& L_q^{\prime\prime}\bigl(
\xi^\star\bigr)\quad\mbox{as } n\to\infty, %
\end{eqnarray*}
almost surely, as we do below; notice that $L_q^{\prime\prime}(\xi
^\star)$
is negative definite and, thus, $R_n(\hat\xi_n,\xi^\star)$ will be
nonsingular,
for large enough $n$, almost surely.

Fix $\varepsilon>0$ and find $\rho>0$ such that
$\|L_q^{\prime\prime}(\xi)-\break L_q^{\prime\prime}(\xi^\star)\|
_\infty<\varepsilon/2$
if $\|\xi-\xi^\star\|_2\le\rho$;
this is possible because $L_q^{\prime\prime}(\cdot)$ is continuous.
Then, observe that $\|\hat\xi_n-\xi^\star\|_2\le\rho$,
for large enough $n$, almost surely, because of (ii). Therefore, we can write
\begin{eqnarray*}
&&
\bigl\|R_n\bigl(\hat\xi_n,\xi^\star
\bigr)-L_q^{\prime\prime}\bigl(\xi^\star\bigr)\bigr\|
_\infty\\
&&\quad< \sup_{\xi:\|\xi-\xi^\star\|_2\le\rho} \bigl\llVert \bar{H}_n
\bigl(y^{(n)}|\xi\bigr)-L_q^{\prime\prime}(\xi)\bigr\rrVert
_\infty+ \frac{\varepsilon}{2}
\end{eqnarray*}
for large enough $n$, almost surely,
where the first term in the right-hand side can be made smaller than
$\varepsilon/2$
by the same uniform Strong Law of Large Numbers invoked above.
The thesis follows, because $\varepsilon$ is arbitrary.
\end{pf}
If $q(\cdot)\in\mathcal{M}$, the asymptotic covariance matrix
$V^\star$
in Theorem~\ref{thmmle} is the inverse of Fisher's information matrix
at $\xi^\star$;
this can be shown through a well-known argument
relying on passing the derivative under the integral sign
(\cite{Ferg1996}, Chapter~18).

Next, in order to study the asymptotic behavior of the marginal likelihood,
we need to introduce the notion of a regular generalized moment prior.
%
\begin{defin}
A generalized moment prior $p^M(\cdot)\propto g(\cdot)p(\cdot)$ on
the open
parameter space $\Xi\subseteq\Re^d$ is regular if the following
assumptions hold:
\begin{enumerate}
\item$p(\cdot)$ is a strictly positive continuous probability density
on $\Xi$
(local prior);
\item$g(\cdot)$ is an infinitely smooth function from $\Xi$ to $\Re_+$,
whose $k$th derivative we denote by $g^{(k)}(\cdot)$;
\item for all $\xi\in\Xi$ the least positive integer $h$ such that
$g^{(2h)}(\xi)\neq0$ (order of the generalized moment prior at $\xi
$) is finite.
\end{enumerate}
It is intended that the normalizing constant $C_g=\int_\Xi g(\xi)
p(\xi)\,d\xi$
be finite, as well as strictly positive, so that $p^M(\cdot)$ is a
proper prior.
\end{defin}
Notice that $g(\xi)=0$ implies $g^\prime(\xi)=0$
and $g^{\prime\prime}(\xi)$ positive semidefinite,
because $g(\cdot)$ is a function to $\Re_+$.
By iterating this argument, we find that\break $g^{(2h-1)}(\xi)=0$ and
$g^{(2h)}(\xi)$
is a positive semidefinite, nonnull, multilinear form on $\Re^{2h}$.

We are now ready to give our main result on the marginal likelihood
of a regular model with a regular generalized moment prior.
%
\begin{theorem}
\label{thmmain}
Let $y^{(n)}=(y_1,\ldots,y_n)$ be data arising under i.i.d. sampling
from an unknown distribution with strictly positive density $q^n(\cdot)$
and $\mathcal{M}=\{f^n(\cdot|\xi),\xi\in\Xi\}$ be a nonsingleton
statistical model
for such data, which we assume to be regular with respect to $q^n(\cdot)$.
Denote by $m^M(y^{(n)})$ the marginal likelihood of $\mathcal{M}$
under a regular generalized moment prior $p^M(\cdot)$. Then:
\begin{longlist}[(ii)]
\item[(i)] if $q(\cdot)\notin\mathcal{M}$,
\[
\log{\frac{m^M(y^{(n)})}{q^n(y^{(n)})}}=-nK^\star+\mathcal{O}_p
\bigl(n^{1/2}\bigr); %
\]
\item[(ii)] if $q(\cdot)\in\mathcal{M}$,
\[
\log{\frac{m^M(y^{(n)})}{q^n(y^{(n)})}}=-\frac{d}{2}\log n-h^\star \log n+
\mathcal{O}_p(1), %
\]
where $h^\star$ is the order of $p^M(\cdot)$ at $\xi^\star$.
\end{longlist}
If $\mathcal{M}$ is a singleton model, which needs no prior,
then \textup{(i)} holds unchanged and \textup{(ii)} holds trivially with $d=0$ and
$h^\star=0$.
\end{theorem}
\begin{pf}
Following \citet{Dawi2011}, we factorize the ratio of the
marginal likelihood
to the unknown sampling distribution as
%
\renewcommand{\theequation}{\arabic{equation}}
\begin{eqnarray}\label{eqfactorization}
\frac{m^M(y^{(n)})}{q^n(y^{(n)})}&=&
\frac{m^M(y^{(n)})}{f^n(y^{(n)}|\hat\xi_n)}\times
\frac{f^n(y^{(n)}|\hat\xi_n)}{f^n(y^{(n)}|\xi^\star)}\nonumber\\[-8pt]\\[-8pt]
&&{}\cdot
\frac{f^n(y^{(n)}|\xi^\star)}{q^n(y^{(n)})}.\nonumber
\end{eqnarray}
We deal with the three factors, which we name $F_1$, $F_2$ and $F_3$,
in reverse order.
Notice that $F_1$ and $F_2$ are (to be considered) identically one for
a singleton model.

The third factor in (\ref{eqfactorization}) is trivially one
if $q(\cdot)\in\mathcal{M}$, because in this case $f(\cdot|\xi
^\star)=q(\cdot)$.
On the other hand, if $q(\cdot)\notin\mathcal{M}$,
its logarithm can be written as
\[
\log{F_3}= \sum_{i=1}^n\log
\frac{f(y_i|\xi^\star)}{q(y_i)}, %
\]
that is, as a sum of i.i.d. random numbers with expectation $-K^\star$.
It follows from the central\vadjust{\goodbreak} limit theorem that
\begin{eqnarray*}
&&
\frac{1}{\sqrt{n}}\bigl(\log{F_3}+nK^\star\bigr)\\
&&\quad
\rightsquigarrow \mathcal{N}_1 \biggl(0,\var_q\log
\frac{q(y_1)}{f(y_1|\xi^\star
)} \biggr)\quad\mbox{as } n\to\infty, %
\end{eqnarray*}
and, thus, we find $\log{F_3}=-nK^\star+\mathcal{O}_p(n^{1/2})$.

The logarithm of the second factor in (\ref{eqfactorization}) can be
written as
\begin{eqnarray*}
\log F_2 &=& - n \int_0^1 (1-u)
\bigl(\xi^\star-\hat\xi_n\bigr)^\top\\
&&\hspace*{28.9pt}{}\cdot
\bar{H}_n\bigl(y^{(n)}|\hat\xi_n+u\bigl(
\xi^\star-\hat\xi_n\bigr)\bigr)\\
&&\hspace*{28.9pt}{}\cdot \bigl(\xi^\star-
\hat\xi_n\bigr)\,du, %
\end{eqnarray*}
using a Taylor expansion with an integral reminder
of the average log-likelihood about $\hat\xi_n$;
remember that $\hat\xi_n$ is a root of the score equation. Like in
the proof of
Theorem~\ref{thmmle}, it can be shown that
\begin{eqnarray*}
&&
\int_0^1 (1-u) \bar{H}_n
\bigl(y^{(n)}|\hat\xi_n+u\bigl(\xi^\star-\hat
\xi_n\bigr)\bigr)\,du\\
&&\quad= \frac{1}{2}L_q^{\prime\prime}
\bigl(\xi^\star\bigr)+o_p(1). %
\end{eqnarray*}
Then, since we know from Theorem~\ref{thmmle}
that\break $n^{1/2}(\hat\xi_n-\xi^\star)=\mathcal{O}_p(1)$, we find
$\log F_2=\mathcal{O}_p(1)$;
this holds regardless of $q(\cdot)\in\mathcal{M}$ or $q(\cdot
)\notin\mathcal{M}$.

The first factor in (\ref{eqfactorization}) can be dealt with by means
of a
Laplace approximation of the marginal likelihood. Specifically,
for all sufficiently small $\rho>0$, the latter can be written as
$m^M(y^{(n)})=I_\rho(y^{(n)})+I^\star_\rho(y^{(n)})$, where
\begin{eqnarray*}
I_\rho\bigl(y^{(n)}\bigr)&=& \int_{\{\xi\in\Xi:\|\xi-\xi^\star\|_2>\rho\}}
f^n\bigl(y^{(n)}|\xi \bigr)p^M(\xi)\,d\xi,
\\
I^\star_\rho\bigl(y^{(n)}\bigr)&=& \int
_{\{\xi\in\Re^d:\|\xi-\xi^\star\|_2\le\rho\}
}f^n\bigl(y^{(n)}|\xi
\bigr)p^M(\xi)\,d\xi.
\end{eqnarray*}
By (iii) of Theorem~\ref{thmmle}, we can find $\delta>0$ such that,
almost surely, for large enough $n$,
$I_\rho(y^{(n)})\le f^n(y^{(n)}|\hat\xi_n)e^{-\delta n}$;
it follows that, by increasing $n$, we can make
$I_\rho(y^{(n)})/\{n^{-h^\star-d/2}f^n(y^{(n)}|\hat\xi_n)\}$
as small as we like. On the other hand,
a Taylor expansion of the average log-likelihood about $\hat\xi_n$
gives us
\begin{eqnarray*}
&&
\frac{I^\star_\rho(y^{(n)})}{f^n(y^{(n)}|\hat\xi_n)}\\
&&\quad= \int_{\{\xi\in\Re^d:\|\xi-\xi^\star\|_2\le\rho\}} \exp \bigl\{-n(\xi-\hat
\xi_n)^\top R_n(\xi,\hat\xi_n) \\
&&\qquad\hspace*{156pt}{}\cdot(
\xi-\hat\xi_n) \bigr\}\\
&&\qquad\hspace*{75pt}{}\cdot p^M(\xi)\,d\xi, %
\end{eqnarray*}
where $R_n(\xi,\hat\xi_n)=
-\int_0^1 (1-u) \bar{H}_n(y^{(n)}|\hat\xi_n+u(\xi-\break\hat\xi_n))\,du$
is the integral reminder.

Now fix $\varepsilon>0$.
Like in the proof of Theorem~\ref{thmmle},
a~suitable choice of $\rho$ makes
$\|R_n(\xi,\hat\xi_n)-\frac{1}{2}K_q^{\prime\prime}(\xi^\star)\|
_\infty$
smaller than $\varepsilon$, for large enough $n$, almost surely.
In this way, we obtain
$J_n(-\varepsilon)\le I^\star_\rho(y^{(n)})/f^n(y^{(n)}|\hat\xi_n)
\le J_n(\varepsilon)$, where
\begin{eqnarray*}
&&
J_n(\varepsilon)\\[-0.5pt]
&&\quad= \int_{\{\xi\in\Re^d:\|\xi-\xi^\star\|_2\le\rho\}} \exp\biggl\{-
\frac{n}{2}(\xi-\hat\xi_n)^\top\\[-0.5pt]
&&\qquad\hspace*{106pt}{}\cdot
\bigl(K_q^{\prime\prime}\bigl(\xi^\star\bigr)-2\varepsilon
d^2I_d\bigr)\\[-0.5pt]
&&\qquad\hspace*{106pt}\hspace*{48.2pt}{}\cdot(\xi-\hat\xi_n)
\biggr\}\\[-0.5pt]
&&\qquad\hspace*{76pt}{}\cdot p^M(\xi)\,d\xi
\end{eqnarray*}
with $I_d$ denoting the $d\times d$ identity matrix.
In the following we deal with $J_n(-\varepsilon)$ implicitly,
by considering $J_n(\varepsilon)$ without assuming $\varepsilon>0$.

Since $p(\xi)=p(\xi^\star)+o(1)$
and $g(\xi)=\frac{1}{(2h^\star)!}\cdot\break g^{(2h^\star)}(\xi^\star)[(\xi
-\xi^\star)^{2h^\star}]
+o(\|\xi-\xi^\star\|_2^{2h^\star})$, as $\xi\to\xi^\star$,
we have
\begin{eqnarray*}
p^M(\xi)&=&\frac{p(\xi^\star)}{C_g(2h^\star)!} g^{(2h^\star)}\bigl(\xi^\star
\bigr)\bigl[\bigl(\xi-\xi^\star\bigr)^{2h^\star}\bigr] \\[-0.5pt]
&&{}+o\bigl(\bigl\|\xi-
\xi^\star\bigr\|_2^{2h^\star}\bigr)\quad \mbox{as $\xi\to\xi
^\star$}, %
\end{eqnarray*}
and a suitable choice of $\rho$ makes
$|J_n(\varepsilon)-\{C_g(2\cdot h^\star)!\}^{-1} p(\xi^\star)J^\star
_n(\varepsilon)|/
\tilde{J}_n(\varepsilon)$ as small as we like,
where
\begin{eqnarray*}
\hspace*{-4pt}&&
J^\star_n(\varepsilon)\\[-0.5pt]
\hspace*{-4pt}&&\quad= \int_{\{\xi\in\Re^d:\|\xi-\xi^\star\|_2\le\rho\}} \exp
\biggl\{-\frac{n}{2}(\xi-\hat\xi_n)^\top\\[-0.5pt]
\hspace*{-4pt}&&\qquad\hspace*{110pt}{}\cdot
\Lambda_\varepsilon (\xi-\hat\xi_n) \biggr\}\\[-0.5pt]
\hspace*{-4pt}&&\qquad\hspace*{80pt}{}\cdot
g^{(2h^\star)}
\bigl(\xi^\star\bigr)\bigl[\bigl(\xi-\xi^\star
\bigr)^{2h^\star}\bigr]\,d\xi %
\end{eqnarray*}
and
\begin{eqnarray*}
&&
\tilde{J}_n(\varepsilon)\\[-0.5pt]
&&\quad= \int_{\{\xi\in\Re^d:\|\xi-\xi^\star\|_2\le\rho\}} \exp
\biggl
\{-\frac{n}{2}(\xi-\hat\xi_n)^\top\\[-0.5pt]
&&\qquad\hspace*{110pt}{}\cdot
\Lambda_\varepsilon (\xi-\hat\xi_n) \biggr\} \\[-0.5pt]
&&\qquad\hspace*{78.3pt}{}\cdot\bigl\|\xi-
\xi^\star\bigr\|_2^{2h^\star}\,d\xi
\end{eqnarray*}
with $\Lambda_\varepsilon=K_q^{\prime\prime}(\xi^\star)-2\varepsilon d^2I_d$.
Both $J^\star_n(\varepsilon)$ and $\tilde{J}_n(\varepsilon)$ are of the form
\begin{eqnarray*}
&&J_n^A(\varepsilon)\\[-0.5pt]
&&\quad= \int_{\{\xi\in\Re^d:\|\xi-\xi^\star\|_2\le\rho\}} \exp
\biggl\{-\frac{n}{2}(\xi-\hat\xi_n)^\top\\[-0.5pt]
&&\qquad\hspace*{110pt}{}\cdot
\Lambda_\varepsilon (\xi-\hat\xi_n) \biggr\} \\[-0.5pt]
&&\qquad\hspace*{78.3pt}{}\cdot A\bigl[\bigl(\xi-
\xi^\star\bigr)^{2h^\star}\bigr]\,d\xi,
\end{eqnarray*}
where $A$ is a positive semidefinite, nonnull, multilinear form on
$\Re^{2h}$.
We consider below the extension of $J_n^A(\varepsilon)$ to $\Re^d$,
which we denote by $\bar{J}_n^A(\varepsilon)$.

By writing $A[(\xi-\xi^\star)^{2h^\star}]=\sum_{i=0}^{2h^\star
}{2h^\star\choose i}
A[(\hat\xi_n-\break\xi^\star)^i(\xi-\hat\xi_n)^{2h^\star-i}]$,
and operating the change of variable $\zeta=n^{1/2}(\xi-\hat\xi
_n)$, we obtain
\begin{eqnarray*}
\bar{J}_n^A(\varepsilon) &=& \int_{\Re^d}
\exp{ \biggl\{-\frac{n}{2}(\xi-\hat\xi_n)^\top
\Lambda_\varepsilon (\xi-\hat\xi_n) \biggr\}} \\
&&\hspace*{17.7pt}{}\cdot A\bigl[\bigl(\xi-
\xi^\star\bigr)^{2h^\star}\bigr]\,d\xi
\\
&=& \sum_{i=0}^{2h^\star}
\pmatrix{2h^\star
\cr
i}\int_{\Re^d} \exp{ \biggl\{-
\frac{1}{2}\zeta^\top\Lambda_\varepsilon\zeta \biggr\}}
\\
&&\hspace*{17.7pt}{}\cdot A
\bigl[\bigl\{n^{1/2}\bigl(\hat\xi_n-\xi^\star\bigr)
\bigr\}^i\zeta^{2h^\star
-i}\bigr]n^{-h^\star-{d/2}}\,d\zeta
\\
&=& \frac{(2\pi)^{d/2}}{|\Lambda_\varepsilon|^
{1/2}}n^{-h^\star-d/2} \vat A\bigl[Z_\varepsilon^{2h^\star}
\bigr]\bigl\{1+\mathcal{O}_p(1)\bigr\},
\end{eqnarray*}
where $Z_\varepsilon$ is a normal random vector with zero mean and precision
matrix $\Lambda_\varepsilon$,
and
$\{1+\mathcal{O}_p(1)\}$ 
is bounded away from zero in probability.
We are now ready to conclude our proof.

Since $|\bar{J}_n^A(\varepsilon)-J_n^A(\varepsilon)|$ is less than
\begin{eqnarray*}
&&\int_{\{\xi\in\Re^d:\|\xi-\hat\xi_n\|_2>\rho/2\}} \exp \biggl\{-\frac{n}{2}(\xi-\hat
\xi_n)^\top\Lambda_\varepsilon (\xi-\hat
\xi_n) \biggr\} \\
&&\hspace*{84.3pt}{}\cdot A\bigl[\bigl(\xi-\xi^\star
\bigr)^{2h^\star}\bigr]\,d\xi, %
\end{eqnarray*}
if $n$ is large enough to have $\|\hat\xi_n-\xi^\star\|_2<\rho/2$,
the same computations carried out above show that\break
$|\bar{J}_n^A(\varepsilon)-J_n^A(\varepsilon)|/n^{-h^\star-d/2}$
is arbitrarily small, for\break large enough~$n$.
Hence, we have $J_n^A(\varepsilon)=(2\pi)^{d/2}\cdot|\Lambda_\varepsilon|^{-1/2}
n^{-h^\star-d/2}\vat A[Z_\varepsilon^{2h^\star}]\{1+\mathcal
{O}_p(1)\}$
and then\break $J_n(\varepsilon)=\{C_g(2h^\star)!\}^{-1}p(\xi^\star)
(2\pi)^{d/2}|\Lambda_\varepsilon|^{-1/2}n^{-h^\star-d/2}\cdot
\vat g^{(2h^\star)}(\xi^\star)[Z_\varepsilon^{2h^\star}]\{1+\mathcal
{O}_p(1)\}$.

Finally, since $\varepsilon$ is arbitrary, it follows that
\begin{eqnarray*}
&&
\frac{m^M(y^{(n)})}{f^n(y^{(n)}|\hat\xi_n)}\\
&&\quad= \frac{p(\xi^\star)}{C_g(2h^\star)!}
\frac{(2\pi)^{d/2}}{|K_q^{\prime\prime}(\xi^\star)|^{1/2}} \\
&&\qquad{}\cdot\vat\bigl
\{g^{(2h^\star)}\bigl(\xi^\star\bigr)\bigl[Z^{2h^\star}\bigr]\bigr
\}n^{-h^\star-{d/2}} \bigl\{1+\mathcal{O}_p(1)\bigr\}, %
\end{eqnarray*}
where $Z$ is a normal random vector with zero mean and precision
matrix $K_q^{\prime\prime}(\xi^\star)$;
this eventually leads to $\log F_1=-h^\star\log n-\frac{d}{2}\log
n+\mathcal{O}_p(1)$
as desired.
\end{pf}
The above theorem also covers local priors,
by letting $g(\xi)\equiv1$, so that $h^\star$ is identically zero;
in this case it essentially returns the result of \citet{Dawi2011}.

We are now in a position to describe the asymptotic behavior of BFs
using a generalized moment prior under the alternative and a local prior
under the null (or comparing to a point null).
%
\begin{cor}
Let $\mathcal{M}_0\subset\mathcal{M}_1$ be two nested models for the
same data
$y^{(n)}=(y_1,\ldots,y_n)$ and assume that both these models are
regular with respect to
all distributions in $\mathcal{M}_1$, with dimensions $d_0<d_1$.
Denote by $\mathrm{BF}^M_{10}(y^{(n)})$ the Bayes factor in favor of $\mathcal{M}_1$
against $\mathcal{M}_0$ using a regular generalized moment prior under
$\mathcal{M}_1$
with order $h$ on the subspace of $\mathcal{M}_1$ corresponding to
$\mathcal{M}_0$.
If $\mathcal{M}_0$ is a nonsingleton model, let it be equipped with a
local prior.
Finally, denote by $q^n(\cdot)$ the actual sampling distribution
and recall that $\mathrm{BF}^M_{01}(y^{(n)})=1/\mathrm{BF}^M_{10}(y^{(n)})$.
Then:
\begin{longlist}[(ii)]
\item[(i)] if $q(\cdot)\in\mathcal{M}_1\setminus\mathcal{M}_0$,
\[
\mathrm{BF}_{01}\bigl(y^{(n)}\bigr)=\exp\bigl\{-nK^\star+
\mathcal{O}_p\bigl(n^{1/2}\bigr)\bigr\}; %
\]
\item[(ii)] if $q(\cdot)\in\mathcal{M}_0$,
\begin{eqnarray*}
&&
\mathrm{BF}_{10}\bigl(y^{(n)}\bigr)\\
&&\quad=\exp \biggl\{-\frac{(d_1-d_0)}{2}
\log n-h\log n+\mathcal{O}_p(1) \biggr\}. %
\end{eqnarray*}
\end{longlist}
\end{cor}
\begin{pf}
This follows directly from Theorem~\ref{thmmain}.
\end{pf}
\end{appendix}

\section*{Acknowledgments}

This work was supported in part by MIUR, Rome,
PRIN2007XECZ7L\_001, and the University of Pavia; it was started
while Consonni and La Rocca were visiting the Southampton Statistical
Sciences Research Institute at the University of Southampton, UK, whose
hospitality and financial support is gratefully acknowledged.

We thank two anonymous referees and an Associate Editor for helping us
improve and extend the scope of our paper.



\end{document}